\documentclass[usenatbib]{mn2e}
\usepackage{amsmath}
\usepackage{natbib}
\usepackage{epsfig}
\usepackage{txfonts}
\usepackage{pict2e}
\usepackage{color}

\newcommand{\Mpc}{{\rm Mpc}}

\newcommand{\expf}[1]{{{\rm e}^{#1}}}

\newcommand{\ion}[2]{{\text{{\sc #1}\,{\sc #2}}}}

\newcommand{\zmu}{{z_{\mu}}}

\newcommand{\nbb}{{n^{\rm pl}}}

\newcommand{\nS}{n_{\rm S}}

\newcommand{\nrun}{n_{\rm run}}

\newcommand{\kD}{k_{\rm D}}

\newcommand{\id}{{\,\rm d}}

\newcommand{\beq}{\begin{equation}}   %

\newcommand{\eeq}{\end{equation}}   %

\newcommand{\beqa}{\begin{eqnarray}}   %

\newcommand{\eeqa}{\end{eqnarray}}   %

\newcommand{\beal}{\begin{align}}

\newcommand{\enal}{\end{align}}

\newcommand{\bspl}{\begin{split}}

\newcommand{\espl}{\end{split}}

\newcommand{\bsub}{\begin{subequations}}

\newcommand{\esub}{\end{subequations}}

\newcommand{\bmulti}{\begin{multline}}   %

\newcommand{\beqm}{\begin{mathletters}}   %

\newcommand{\eeqm}{\end{mathletters}}   %

\newcommand{\pot}[2]{#1 \times 10^{#2}}

\usepackage{hyperref}
\usepackage{grffile}
\usepackage{graphics}
\usepackage{booktabs}

\title[Distortion constraints]
{Distinguishing different scenarios of early energy release with spectral distortions of the cosmic microwave background}

\author[Chluba]{
 J.~Chluba$^{1}$\thanks{E-mail: jchluba@pha.jhu.edu}
 \\
$^{1}$ Department of Physics and Astronomy, Johns Hopkins University, Bloomberg Center 435, 
3400 N. Charles St., Baltimore, MD 21218, USA
}

\begin{document}

\date{{Accepted 2013 September 11. Received 2013 April 22}}

\maketitle

\begin{abstract}
Deviations of the cosmic microwave background (CMB) frequency spectrum from a pure blackbody tell an exciting story about the thermal history of our Universe. In this paper, we illustrate how well future CMB measurements might decipher this tale, envisioning a {\it PIXIE}-like spectrometer, which could improve the distortion constraints obtained with {\it COBE}/{\rm FIRAS} some 20 years ago by at least three orders of magnitude. This opens a large discovery space, offering deep insights to particle and early-universe physics, opportunities that no longer should be left unexplored. Specifically, we consider scenarios with annihilating and decaying relic particles, as well as signatures from the dissipation of primordial small-scale power. {\it PIXIE} can potentially rule out different early-universe scenarios, and moreover will allow unambiguous detections in many of the considered cases, as we demonstrate here. We also discuss slightly more futuristic experiments, with several times improved sensitivities, to highlight the large potential of this new window to the pre-recombination universe.
\end{abstract}


\section{Introduction}
\label{sec:Intro}
Cosmology is now a precise scientific discipline, with detailed theoretical models that fit a wealth of very accurate measurements. Of the many cosmological data sets, the cosmic microwave background (CMB) {\it anisotropies} provide the most stringent and robust constraints to theoretical models, allowing us to address fundamental questions about inflation, the nature of dark matter and dark energy, and particle physics \citep{Smoot1992, WMAP_params, Planck2013params}.
But the CMB holds another, complementary piece of information: its {\it frequency spectrum}. Since the measurements with {\it COBE}/{\rm FIRAS} the average CMB spectrum is known to be extremely close to a perfect blackbody, with possible spectral distortions limited to $\Delta I_\nu/I_\nu \lesssim \pot{\rm few}{-5}$ \citep{Mather1994, Fixsen1996}. Although thus far no CMB distortion was detected, this impressive measurement already places very tight constraints on the thermal history of our Universe, ruling out cosmologies with extended periods of significant energy release, disturbing the equilibrium between matter and radiation.

More than 20 years have passed since the launch of {\it COBE}, and from the technological point of view already today it should be possible to improve the absolute spectral sensitivity by at least three orders of magnitude \citep{Kogut2011PIXIE}. 
This opens a new window to the early Universe, on one hand allowing us to directly probe processes that are present within the standard cosmological paradigm, and on the other hand also opening up a huge discovery space to unexplored non-standard physics.
It is therefore time to ask what exactly one might hope to learn from future measurements of the CMB spectrum and how well could the information be extracted.

The CMB spectrum constrains energy release occurring at redshifts $z \lesssim \zmu \simeq \pot{2}{6}$.  Above $\zmu$, when the Universe was only a few months old, the cosmological thermalization process was extremely efficient, exponentially suppressing any distortion the earlier the energy was liberated \citep{Zeldovich1969, Sunyaev1970mu, Illarionov1975, Illarionov1975b, Danese1977, Burigana1991, Hu1993, Chluba2005, Chluba2011therm, Khatri2012b}. At slightly lower redshift, the CMB spectrum becomes vulnerable, and any energy injection \textit{initially} gives rise to a Compton $y$-distortion in the CMB blackbody, essentially by an early-universe analogue of the Sunyaev-Zeldovich effect. At $\pot{5}{4}\lesssim z\lesssim \zmu$, the distortion rapidly evolves towards a chemical potential or $\mu$-type distortion, due to efficient redistribution of photons over frequency, while for energy release occurring at $z\lesssim \pot{5}{4}$, it keeps the initial shape of a $y$-distortion.

This is, however, not the end of the story. First of all, there is no sharp boundary at $z\simeq \zmu$, and the more sensitive a experiment is, the deeper can one in principle probe beyond the {\it distortion visibility function}, $\mathcal{J}(z)\simeq \exp\left[-(z/\zmu)^{5/2}\right]$, defined by the efficiency of photon production, although the challenge does grow exponentially.
Secondly, the transition from $\mu$- to $y$-distortion occurs more gradually, with the exact shape of the distortion at the intermediate stages depending directly on the energy-release history at $10^4 \lesssim z\lesssim \pot{3}{5}$. 
Signals produced mainly during this epoch were discussed in \citet[][see Fig.~15 and 19]{Chluba2011therm}, showing that the total distortion is not simply given as a superposition of pure $\mu$- and $y$-distortions. 
Also, the numerical studies of \citet{Burigana1991} and \citet{Hu1995PhD} mentioned this aspect of the problem, which more recently, was clearly demonstrated by \citet{Khatri2012mix} and \citet{Chluba2013Green}.
The small ($\simeq 10\%-30\%$) residual (non-$\mu$/non-$y$) provides additional leverage for distinguishing different energy-release scenarios in the future, although only a small fraction ($\simeq 10\%-20\%$) of the liberated energy is carried by this signal, making it a correction to the main mix of $\mu$- and $y$-distortions \citep{Chluba2013Green}.

A large number of astrophysical or cosmological processes at $z\lesssim \zmu$ exist, leading to predictions of observable distortions:

\begin{itemize}

\item[$\bullet$] {\it Reionization and structure formation:} the first sources of radiation during reionization \citep{Hu1994pert}, supernova feedback \citep{Oh2003} and structure formation shocks \citep{Sunyaev1972b, Cen1999, Refregier2000, Miniati2000} heat the intergalactic medium at low redshifts ($z\lesssim 10$), leading to partial up-scattering of CMB photons, causing a Compton $y$-distortion. 
The distortion is expected to reach $\Delta I_\nu/I_\nu \simeq 10^{-7}-10^{-6}$ and thus could be measured at $\simeq 100\,\sigma$ using present-day technology \citep{Kogut2011PIXIE}, teaching us about the average temperature of the intergalactic medium \citep[e.g.,][]{Zhang2004} and promising a way to find the missing baryons in the local Universe, which otherwise are hard to observe \citep{Cen1999}.

\vspace{2mm}

\item[$\bullet$] {\it Inflation:} the Silk-damping of small-scale perturbations gives rise to both $\mu$- and $y$-type distortions \citep{Sunyaev1970diss,Daly1991,Barrow1991,Hu1994}, which directly depend on the shape and amplitude of the primordial power spectrum at scales $0.1\,{\rm kpc}\lesssim \lambda \lesssim 1\,{\rm Mpc}$ \citep{Chluba2012,Khatri2012short2x2}. This allows constraining the trajectory of the inflaton at stages unexplored by CMB anisotropies and other ongoing or planned experiments \citep{Chluba2012inflaton}. 
The distortion in principle is also sensitive to the difference between adiabatic and isocurvature perturbations \citep{Barrow1991,Hu1994isocurv, Dent2012, Chluba2013iso}, as well as primordial non-Gaussianity in the ultra squeezed-limit, leading to a spatially varying spectral signal that correlates with CMB temperature anisotropies at large scales \citep{Pajer2012, Ganc2012, Biagetti2013}.

\vspace{2mm}

\item[$\bullet$] {\it Decaying or annihilating relics:} measurements of the CMB spectrum also have the potential to constrain decaying and annihilating particles in the pre-recombination epoch \citep{Hu1993b, McDonald2001, Chluba2010a, Chluba2011therm}
This is especially interesting for decaying particles with lifetimes $t_{\rm X} \simeq \pot{3}{8}\,{\rm sec}-\pot{2}{11}\,{\rm sec}$, because the shape of the distortion encodes when it decayed \citep{Chluba2011therm}.
Spectral distortions thus provide a probe of particle physics in the early Universe that is complementary to constraints from light-element abundances or CMB anisotropies (see Sect.~\ref{sec:annihil} and \ref{sec:decay} for more details).

\vspace{2mm}

\item[$\bullet$] {\it Cosmological recombination radiation:} the cosmological recombination process of hydrogen and helium introduces distortions \citep{Zeldovich68, Peebles68, Dubrovich1975} at high redshifts ($z\simeq 10^3-10^4$), corresponding to $\simeq 260\,{\rm kyr}$ (\ion{H}{i}), $\simeq 130\,{\rm kyr}$  (\ion{He}{i}) and $\simeq 18\,{\rm kyr}$ (\ion{He}{ii}) after the big bang \citep{Jose2006, Chluba2006b, Jose2008}. The overall signal is very small ($\Delta I_\nu/I_\nu \simeq 10^{-9}$ close to the maximum of the CMB blackbody), but it has a unique frequency dependence which opens an independent path for determination of cosmological parameters (like the baryon density and {\it pre-stellar} helium abundance) and direct measurements of the recombination dynamics, probing the Universe at stages well before the last scattering surface \citep{Chluba2008T0, Sunyaev2009}.

\vspace{2mm}

\item[$\bullet$] {\it Cooling of matter:} the adiabatic cooling of ordinary matter continuously extracts energy from the CMB photon bath by Compton cooling leading to a small but indisputable distortion that directly depends on the baryon density and is characterized by {\it negative} $\mu$- and $y$-parameters at the level of $\simeq \pot{\rm few}{-9}$ \citep{Chluba2005, Chluba2011therm, Khatri2011BE}.

\end{itemize}

\noindent
All these examples demonstrate that the CMB spectrum provides a rich and unique source of complementary information about the early Universe, with the certainty for a detection of spectral distortions at a level within reach of present-day and future instrumentation.
The CMB spectrum could also allow placing interesting constraints on the power spectrum of small-scale magnetic fields \citep{Jedamzik2000}, primordial black holes \citep{Carr2010}, cosmic strings \citep{Ostriker1987, Tashiro2012, Tashiro2012b} and other new physics \citep{Lochan2012, Bull2013} to mention a few more exotic examples.
Deciphering all these signals will be a big challenge for the future, but it holds the potential for new discoveries, providing additional, independent constraints on unexplored processes that otherwise might remain a secret of our Universe.

In this paper, we investigate how well future measurement of the CMB spectrum might be able to constrain and {\it distinguish} different mechanisms. In addition to the huge $y$-distortion introduced at low redshifts, we consider scenarios of decaying and annihilating relic particles and the dissipation of small-scale acoustics modes as representative examples. These mechanisms can all potentially lead to large distortions, compatible with existing limits, and thus emphasize the potential of future CMB spectrum measurements.
We fix the background cosmology and only vary parameters related to the distortions. This also means that the signal
caused by the adiabatic cooling of ordinary matter can be predicted with very high precision, and thus is taken out.
The signatures from the recombination epoch are a few times below detection limit of a {\it PIXIE}-type experiment, and we leave a more in depth discussion to future work.

Our estimates demonstrate that {\it PIXIE} can not only rule out different energy-release scenarios, but will also allow unambiguous detections for many of the considered examples. However, the constraints remain model dependent and degeneracies between different scenarios exist, which can only be broken at higher sensitivity. We, therefore, also discuss slightly more futuristic experiments, with improved sensitivities, highlighting the large discovery potential of this unexplored window to the early Universe.

\section{Mock spectral distortion data}
\label{sec:mock}
Energy release in the early Universe causes spectral distortions of the CMB, with the shape of the distortion depending on how strongly the thermal history is affected.
For a specific scenario one can accurately predict the spectral distortion at different frequencies \citep[e.g., see][]{Chluba2011therm, Chluba2013Green}. 
To answer how well different scenarios can be constrained, we have to produce mock spectral distortion data. We envision a {\it PIXIE}-like experiment, with many equidistant channels over a wide range of frequencies ($30\,{\rm GHz} \lesssim \nu \lesssim 6\,{\rm THz}$). As a first step, we shall assume that the measurement is only limited by uncorrelated instrumental noise. Foregrounds due to dust and synchrotron emission were removed to a level below this sensitivity, making use of high-frequency channels ($\nu\gtrsim 1\,{\rm THz}$) and spatial templates [e.g., obtained with {\it Planck} \citep{Planck2013components}].
It seems that these requirements can be achieved in the future, with realistic error bars $\Delta I_\nu\simeq \pot{5}{-26}\,{\rm W\,m^{-2}\,s^{-1}\,Hz^{-1}\,sr^{-1}}$ per $\Delta \nu \simeq 15\,{\rm GHz}$ channel over a $t \simeq 0.5\,{\rm yr}$ measurement period\footnote{Currently, {\it PIXIE} is meant to devote only $\simeq 25\%$ of the 2 years observing time to measurements of the CMB spectrum (Kogut, priv. com.).} \citep{Kogut2011PIXIE}.

The spectral distortion signal from early energy release is mainly important in the $30\,{\rm GHz} \lesssim \nu \lesssim 1\,{\rm THz}$ channels. Unless stated otherwise, we assume that all channels in this frequency range can be used to constrain the thermal history. 
We furthermore assume that the channels are independent and described by a top-hat filter, $W_i(\nu)$, centered at frequency $\nu_i$. 

To accelerate the parameter estimation process, we compute the distortions for different scenarios using a Green's function of the cosmological thermalization problem \citep{Chluba2013Green}:
\beq\label{eq:final_dist}
\Delta I_\nu(z=0)
=\int G_{\rm th}(\nu, z', 0) \, \frac{\id (Q/\rho_\gamma)}{\id z'} \id z'.
\eeq
Here $\id (Q/\rho_\gamma)/\id z$ is the effective heating rate and $\rho_\gamma\propto T_0^4$ the energy density of the undistorted CMB. We also assume that the background cosmology is fixed to $Y_{\rm p}=0.24$, $\Omega_{\rm m}=0.26$, $\Omega_{\rm b}=0.044$, $\Omega_{\Lambda}=0.74$, $\Omega_{\rm k}=0$, $h=0.71$, and $N_{\rm eff}=3.046$. Since we only need the average signal in specific band, we first compute the averages of $G_{i, \rm th}(z, 0)=\int W_i(\nu)\,G_{\rm th}(\nu, z, 0)\id \nu$ in channel $i$. This further reduces the total computational burden.

To compute constraints on the different scenarios, we use a Markov Chain Monte Carlo method (MCMC), adapting routines that were developed as part of {\sc SZpack} \citep{ChlubaSZpack, Chluba2012moments} and that are based on the {\sc Python} packages of \citet{Foreman2012}. Alternatively, one can use simple Fisher forecasts; however, the parameter space becomes very large  and degenerate close to the detection limit, so that we chose to follow an MCMC approach throughout.
We place very conservative priors on the parameters and usually ran chains with $\simeq 10^5$ samples.
Thanks to the adopted Green's function approach, this is possible on a standard laptop in only a few seconds to minutes.
The developed tools are now part of {\sc CosmoTherm}\footnote{{\sc CosmoTherm} is available at \url{www.Chluba.de/CosmoTherm}.} \citep{Chluba2011therm}.

\subsection{Late-time $y$-distortion}
One of the dominant distortion signals is caused by the heating of matter during the reionization epoch, introducing a $y$-distortion,
\beal
\label{eq:DI_SZ}
\Delta I_\nu/y&=\frac{2h\nu^3}{c^2} Y_{\rm SZ}(x) 
=\frac{2h\nu^3}{c^2}\frac{x\expf{x}}{(\expf{x}-1)^2}\left[x\coth(x/2)-4\right],
\end{align}
with Compton parameter, $y\simeq \pot{\rm few}{-7}$ \citep{Hu1994pert}.  We shall use $y_{\rm re}=\pot{4}{-7}$ as a fiducial value, assuming that $y\simeq \pot{2}{-7}$ is caused by Compton scattering of CMB photons by thermal electrons, while another $y\simeq \pot{2}{-7}$ is related to other heating mechanisms mentioned above. This is rather conservative, since the low-redshift signal in principle can exceed the level of $y\simeq 10^{-6}$ \citep[e.g.,][]{Refregier2000, Oh2003}. This will make a detection of the late-time $y$-distortion only easier, and in terms of the error budget (as foreground for the smaller primordial signal) our numbers should produce pretty reliable estimates.

\subsection{Parametrization of the energy-release mechanisms}
\label{sec:parametrization}
The following parametrization for the energy-release mechanisms shall be considered \citep[see][for more details]{Chluba2011therm}:
\begin{itemize}

\item[$\circ$] decaying particle: $\frac{\id (Q/\rho_\gamma)}{\id z}=\frac{f_{\rm X} \, \Gamma_{\rm X}\, N_{\rm H}}{H\,\rho_\gamma\,(1+z)} \, \expf{-\Gamma_{\rm X} t}\propto z^{-4}\,\expf{-(z_{\rm X}/z)^{2}}$

\vspace{2mm}

\item[$\circ$] s-wave annihilation: $\frac{\id (Q/\rho_\gamma)}{\id z}=\frac{N_{\rm H} (1+z)^2}{H\,\rho_\gamma} f_{\rm ann} \propto z^{-1}$

\vspace{2mm}

\item[$\circ$] p-wave annihilation: 
\beal
\frac{\id (Q/\rho_\gamma)}{\id z} =
\begin{cases}
&\frac{N_{\rm H} (1+z)^3}{H\,\rho_\gamma} f_{\rm ann} \propto {\rm const} 
\,\,\,\quad{\rm for}\;\left<\sigma \rm v\right>\propto (1+z)
\\[2mm]
&\frac{N_{\rm H} (1+z)^4}{H\,\rho_\gamma} f_{\rm ann} \propto (1+z) 
\quad{\rm for}\;\left<\sigma \rm v\right>\propto (1+z)^2
\end{cases}
\end{align}

\vspace{2mm}

\item[$\circ$] Silk-damping: $\frac{\id (Q/\rho_\gamma)}{\id z} \propto z^{-1}$ (for scale-invariant primordial power spectrum of curvature perturbations)

\end{itemize}
where $N_{\rm H}(z)\simeq \pot{1.9}{-7}\,(1+z)^3$ denotes the number density of hydrogen nuclei and $H(z)$ the Hubble parameter.

The model parameters for the decaying particle scenario are, $f_{\rm X}$ (which is related to the mass, $m_{\rm X}$ and abundance, $N_{\rm X}$, of the relic particle and the decay channels), and the lifetime of the decaying particle, $t_{\rm X}=\Gamma^{-1}_{\rm X}$.
The two annihilation scenarios only depend on the annihilation efficiency, $f_{\rm ann}$ \citep[see][for more explanation]{Chluba2011therm}, which again is related to the mass and abundance of the relic particle and the decay channels. We furthermore assumed different cross-sections, representing s-wave $[\left<\sigma \rm v\right>\simeq {\rm const}]$ and p-wave annihilation $[\left<\sigma \rm v\right> \propto {\rm v}^2]$ \citep[see][for related discussion]{McDonald2001}. The first p-wave scenario corresponds to a Majorana particle which either is still relativistic after freeze out [e.g., a sterile neutrino with low abundance \citep{Ho2013}], or shows $1/\rm v$ Sommerfeld-enhanced annihilation cross-section \citep[e.g., see][]{Chen2013}, i.e. $\left<\sigma \rm v\right>\propto (1+z)$. Thus, for this scenario a factor $kT_{\rm X,0}/m_{\rm X}c^2$ was absorbed in the definition of $f_{\rm ann}$. For a non-relativistic Majorana particle (second p-wave annihilation scenario) the cross-section scales even faster with redshift, $\left<\sigma \rm v\right>\propto (1+z)^2$, causing practically no energy release at late times.

For both the annihilation and decaying particle scenarios, we furthermore, have to multiply the total  matter heating rate given above by the energy branching ratio, $g_{\rm h}(z)$, for which we follow \citet{Chen2004} and \citet{Chluba2010}. This factor simply takes into account that at different redshifts not all the released energy causes heating: at $z\gtrsim 10^4$ one has $g_{\rm h}(z)\simeq 1$, while after recombination $g_{\rm h}(z)\ll 1$. The exact redshift scaling depends on the recombination history \citep[we use {\sc CosmoRec},][]{Chluba2010b} and how efficiently the decay products transfer their energy to the medium. More sophisticated calculations for the heating efficiencies based on different  particle models can be found in, e.g., \citet{Slatyer2009} and \citet{Valdes2010}, but for the purpose of this paper, the approximation mentioned above will suffice.

To describe the energy release caused by the dissipation of acoustic modes at small scales we follow \citet{Chluba2013iso}, including only adiabatic modes with 
\beal
\label{eq:Q_ac_eff}
\frac{\id (Q/\rho_\gamma)}{\id z}
& \approx 
2 D^2  \int^\infty_{k_{\rm cut}} \mathcal{P}_i(k) \, \partial_z e^{-2k^2/\kD^2} \id\ln k,
\end{align}
where $\mathcal{P}_i(k)\equiv A_i \,(k/k_0)^{\nS-1+\frac{1}{2} n_{\rm run} \ln(k/k_0)}$ is the small-scale power spectrum of curvature perturbations, $\kD(z)$ the dissipation scales and $D^2\simeq 0.81$ the heating efficiency for adiabatic modes (assuming $N_{\rm eft}=3.046$). 
The distortion does depend on the type of initial conditions (adiabatic versus isocurvature); however, as shown by \citet{Chluba2013iso}, constraints can only be derived in a model dependent way and thus discussion of adiabatic modes sweeps the whole possible parameter space (modulo overall efficiency factors and changes in the shape of the primordial power spectrum to accommodate for the differences between the perturbation modes).
As cutoff scales we choose $k_{\rm cut}\simeq 0.12\,\Mpc^{-1}$, which reproduces the heating rate caused by mode dissipation pretty well, even around the recombination epoch.
To minimize the time spent on numerical integration, given the power spectrum parameters, we tabulate the heating rate prior to the computation of the distortions.

\subsection{Shift in the monopole temperature}
The CMB monopole temperature is known with extraordinary accuracy, $T_0=2.7260 \pm 0.0013\,{\rm K}$ \citep{Fixsen2009}. However, the level of precision that might be achievable with a {\it PIXIE}-type experiment will dwarf this measurement.
In the thermalization calculations, we assumed $T_0\equiv 2.726\,{\rm K}$. The error that is introduced by this assumption is at most $\simeq 0.05\%$ relative to the predicted distortion, a margin one can comfortably live with.
We must, however, take the possible shift in the temperature of the reference blackbody into account. At first order in $\Delta\equiv \Delta T/T$, this is just a temperature shift term, but even the second order correction [a $y$-distortion \citep{Chluba2004}] has to be considered, since the error in the precise values of $T_0$ itself corresponds to $y\simeq (\pot{5}{-4})^2/2\simeq 10^{-7}$.
For the parameter estimation problem we thus add
\beal
\label{eq:DI_temp}
\Delta I_\nu&=\frac{2h\nu^3}{c^2} \left[\nbb(x)-\nbb(x/[1+\Delta])\right]
\nonumber \\
&= \frac{2h\nu^3}{c^2} \left[G(x)\Delta [1+\Delta]+Y_{\rm SZ}(x)\,\frac{\Delta^2}{2} \right]+\mathcal{O}(\Delta^3)
\end{align}
where $\nbb(x)=[\expf{x}-1]^{-1}$, $G(x)=x\,\expf{x}/[\expf{x}-1]^2$ describes a temperature shift, and $Y_{\rm SZ}(x)$ a $y$-distortion term with $x=h\nu/k T_0$. For our simulations, we use $\Delta = \pot{1.2}{-4}$ as fiducial value (this is just made up) and then show how well one will be able to constrain it, assuming a Gaussian prior with width $\simeq \pot{5}{-4}$ around it. This is very conservative, since a {\it PIXIE}-type experiment could determine the CMB monopole temperature with $1\sigma$-precision of $\Delta T\simeq 3\,{\rm nK}$ (see below).
We also assume that the effects caused by the superposition of blackbodies related to the motion-induced CMB dipole is taken out. This leads to a $y$-distortion quadrupole with $y\simeq \pot{2.563}{-7}$, and a shift of the CMB monopole temperature by $\Delta T\simeq 0.699\,\mu{\rm K}$ \citep{Chluba2004, Chluba2011therm, Sunyaev2013dipole}.

To accelerate the parameter estimation, we again first compute the averages over the frequency filters. We also use this procedure for the annihilation and decaying particle scenarios, since the parameter dependence is sufficiently simple. For the energy release caused by the dissipation of acoustic modes, we explicitly integrate the Green's function, but we tabulate the energy-release history once the power spectrum parameters are chosen.

\begin{figure}
\centering
\hspace{-3mm}\includegraphics[width=1.02\columnwidth]{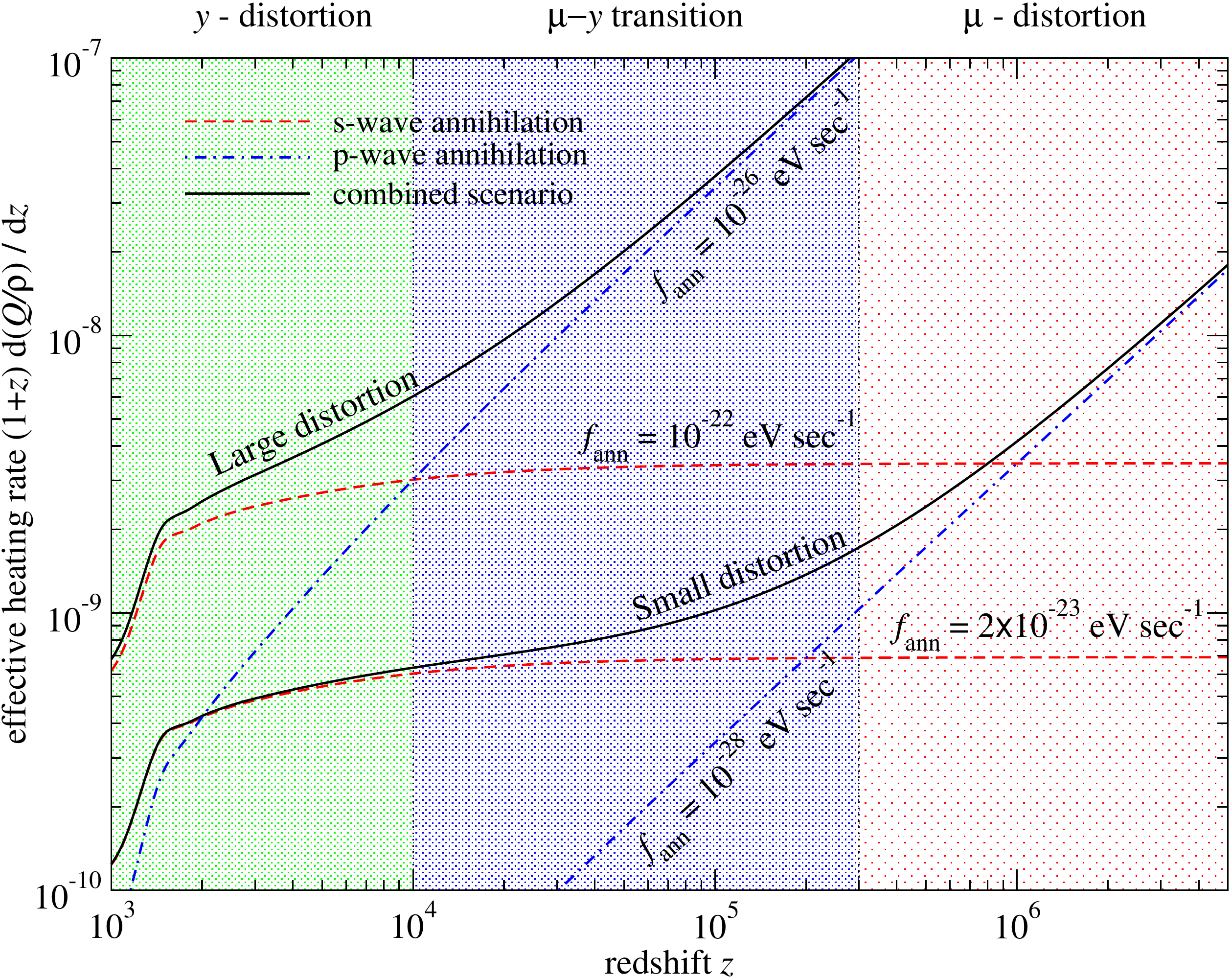}
\\[4mm]
\hspace{-3mm}\includegraphics[width=1.02\columnwidth]{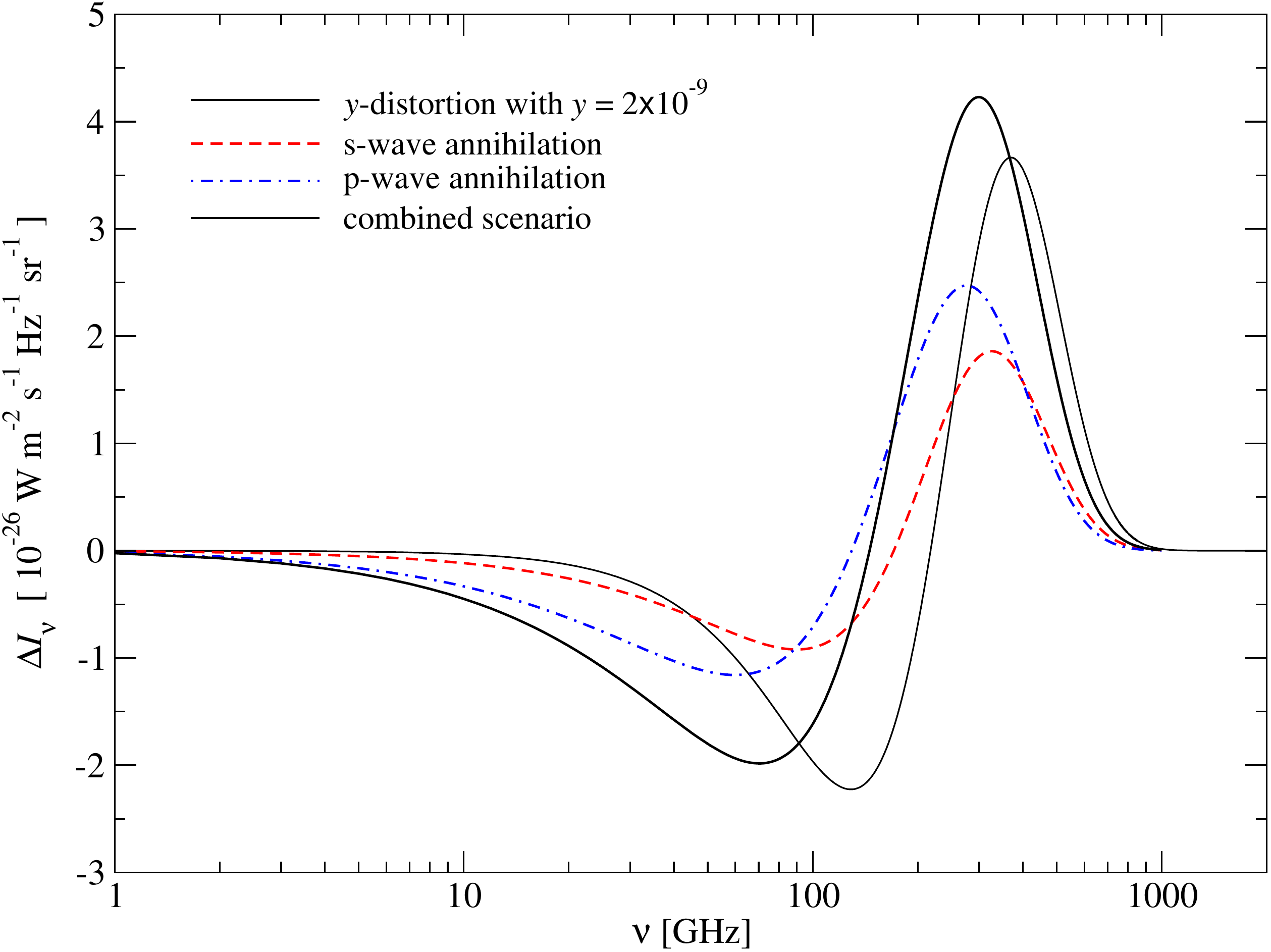}
\caption{Different s- and p-wave annihilation scenarios discussed in Sect.~\ref{sec:annihil}. The upper panel shows the energy-release rate for all cases, while the lower panel only illustrates the spectral signal for the small distortion scenario. For comparison, we show a $y$-distortion of $y=\pot{2}{-9}$, which for {\it PIXIE} sensitivity, $\Delta I_\nu\simeq \pot{5}{-26}\,{\rm W\,m^{-2}\,s^{-1}\,Hz^{-1}\,sr^{-1}}$, should be detectable at the $1\sigma$-level. An unambiguous detection of the signal from the small distortion scenarios will be challenging even at $\simeq 4$ times the sensitivity of {\it PIXIE}, but the large distortion scenarios can be tightly constrained. The amplitude of the distortion signal is directly proportional to the annihilation efficiency, while the shape just depends on the temperature/velocity dependence of the annihilation cross-section (s-wave versus p-wave)}
\label{fig:DM_scenarios}
\end{figure}

\section{Annihilating particle scenarios}
\label{sec:annihil}
As first simple scenario, we consider an annihilating particle with p-wave annihilation cross-section $\left<\sigma v\right> \propto (1+z)$ [see Sect.~\ref{sec:parametrization} for explanation]. 
Constraints on this case can be derived from BBN (due to the sensitivity of the light-element abundances on the baryon-to-photon ratio, $\eta$), implying that the total amount of energy release at that epoch cannot exceed $\Delta \rho_\gamma/\rho_\gamma \simeq 5\%$ \citep{Steigman2007}. This places a bound $f_{\rm ann}\lesssim \pot{4}{-24}\,{\rm eV \,sec^{-1}}$ on the annihilation efficiency.
{\it COBE}/{\rm FIRAS} constraints are a factor of $\simeq 3$ more stringent, implying $f_{\rm ann}\lesssim \pot{1.5}{-24}\,{\rm eV \,sec^{-1}}$ from $|\mu|\lesssim \pot{9}{-5}$ (95\% c.l.). 
Another tight limit derives from measurement of the CMB temperature and polarization anisotropies, corresponding to $f_{\rm ann}\lesssim 10^{-26}\,{\rm eV \,sec^{-1}}$, as we argue below.
Still, this suggests that in principle large energy release can be accommodated for this scenario, without violating existing constraints. Due to the redshift dependence of the heating rate, most energy is liberated during the $\mu$-era (see Fig.~\ref{fig:DM_scenarios}), and hence the distortion should be easily distinguishable from the large $y$-distortion created at low redshifts.

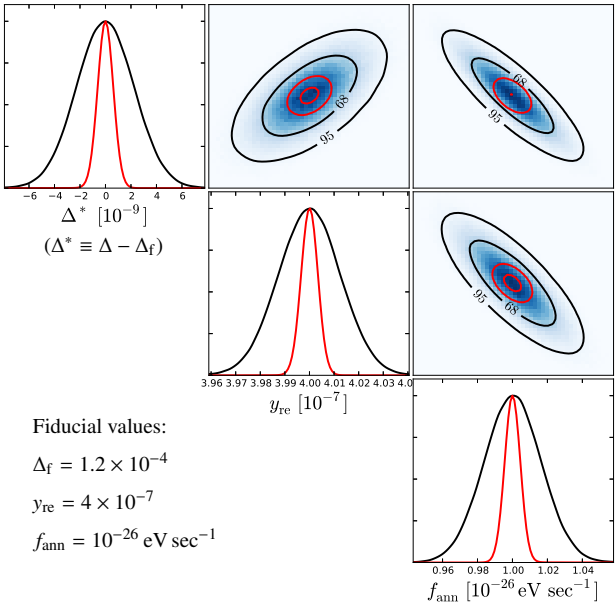
\begin{figure}
\centering
\setlength{\unitlength}{0.05\columnwidth}
  \begin{picture}(20,20)
    \put(0,0){\includegraphics[width=\columnwidth]{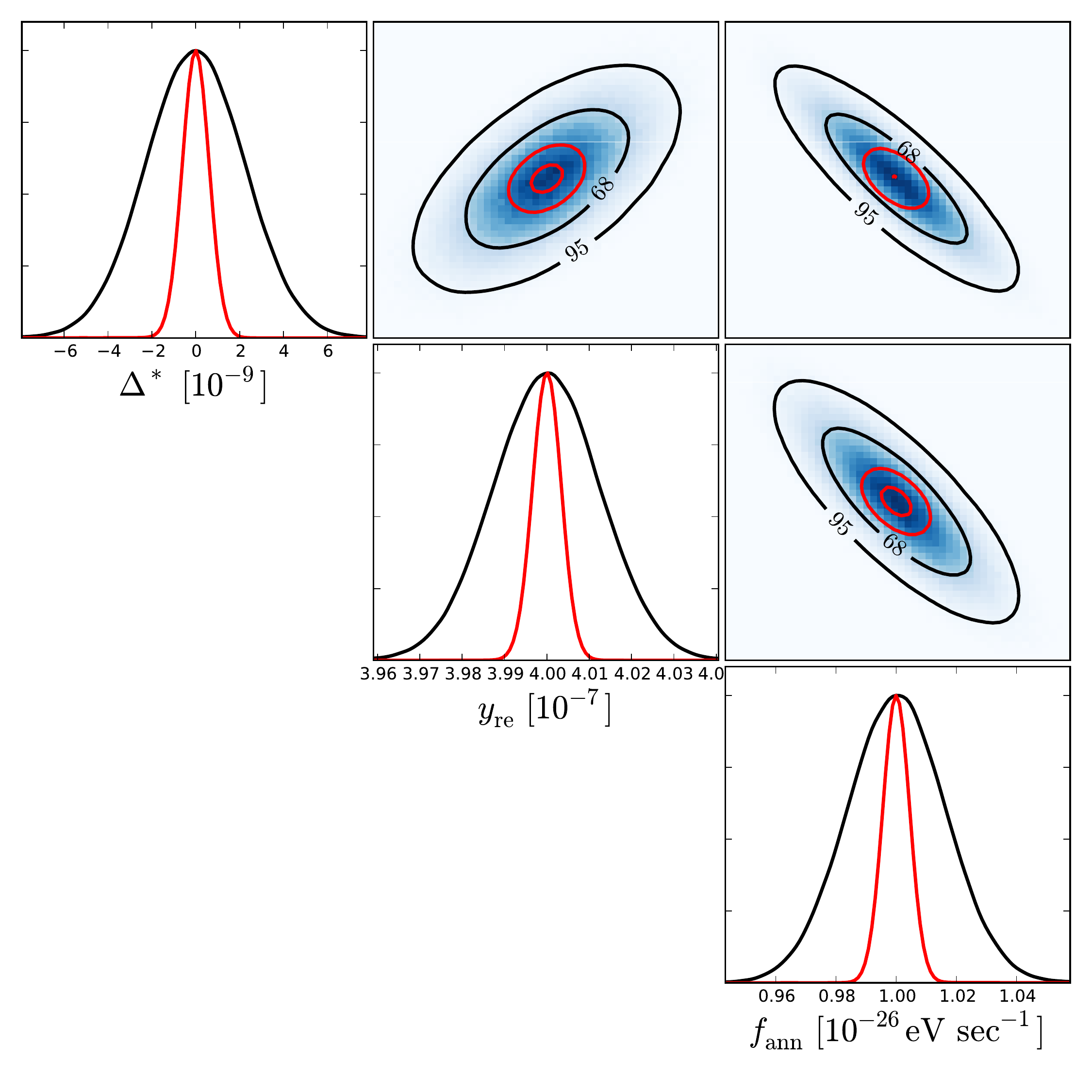}}
    \put(1.65,11.8){$(\Delta^\ast\equiv\Delta-\Delta_{\rm f})$}
    \put(1.3,6.1){Fiducial values:}
    \put(1.3,4.9){$\Delta_{\rm f}=\pot{1.2}{-4}$}
    \put(1.3,3.7){$y_{\rm re}=\pot{4}{-7}$}
    \put(1.3,2.5){$f_{\rm ann}=10^{-26}\,{\rm eV \, sec^{-1}}$}
  \end{picture}
\caption{Large p-wave annihilation scenario. The solid black lines show the constraint for {\it PIXIE} sensitivity, while the red curves are for 4 times higher sensitivity. The contours show $68\%$ and $95\%$ confidence levels. The shaded regions illustrate the shape of the projected 2D probability distribution function for {\it PIXIE} sensitivity only. The marginalized distributions were all normalized to unity at the maximum.}
\label{fig:pwave_DM}
\end{figure}

In Fig.~\ref{fig:pwave_DM}, we show the projected constraints for a {\it PIXIE}-type experiment in a large p-wave annihilation cross-section scenario, with $\Delta\rho_{\gamma}/\rho_\gamma\simeq \pot{6.2}{-7}$ going into the distortion. 
For $y_{\rm re}$ we assumed a flat prior over the interval $y_{\rm re}\in [0, \pot{1.5}{-5}]$, while we sampled $f_{\rm ann}$ uniformly between 0 and 100 times the input value.
Although this prior was rather wide, the MCMC computation converged very rapidly, using about $10^{5}$ samples.
A {\it PIXIE}-type experiment will easily distinguish the associated distortion from the reionization signal, measuring $f_{\rm ann}$ with $\simeq 2\%$ ($1\sigma$-error) precision. 
Since the signal is directly proportional to $f_{\rm ann}$, we find
\beal
\label{eq:error_p_wave}
\frac{\Delta f_{\rm ann, p}}{f_{\rm ann,p}}
& \approx 
2\%\,\left[\frac{f_{\rm ann,p}}{10^{-26}\,{\rm eV \,sec^{-1}}}\right]^{-1}\,\left[\frac{\Delta I_\nu}{\Delta I^{\rm PIXIE}_\nu}\right]^{-1}
\end{align}
for the error, where $\Delta I^{\rm PIXIE}_\nu\simeq \pot{5}{-26}\,{\rm W\,m^{-2}\,s^{-1}\,Hz^{-1}\,sr^{-1}}$ denotes {\it PIXIE}'s sensitivity (we confirmed this statement numerically). The rough $1\sigma$-detection limit of {\it PIXIE} therefore is $f_{\rm ann, p}\simeq \pot{2}{-28}\,{\rm eV \,sec^{-1}}$.
Increasing the sensitivity 2 or 4 times might be within reach, e.g., by extending the total time spent on spectral distortion measurements or by slightly improving the instrument. As our results show, this would further tighten possible limits on this scenario, allowing us to constrain Majorana particles annihilating into lighter fermions \citep{Goldberg1983}.

Figure~\ref{fig:pwave_DM} also shows that the monopole temperature and reionization $y$-parameter could be measured with impressive accuracy, corresponding to $\Delta T\simeq 3\,{\rm nK}$ and $\Delta y_{\rm re}/y_{\rm re}\lesssim 1\%$. 
Both $\Delta$ and $y_{\rm re}$ are anti correlated with $f_{\rm ann}$: although the annihilation distortion signal does not include any pure temperature shift contribution, it is not fully orthogonal to the signal related to $\Delta$ [see. Eq.~\eqref{eq:DI_temp}]. Similarly, every annihilation is associated with some late energy release ($z\lesssim 10^4$), during the $y$-era, and thus boosted annihilation efficiency leaves less room for contribution to $y$ from after recombination and during reionization, explaining the behavior.

Assuming a relic particle with $f_{\rm ann, p}\simeq10^{-28}\,{\rm eV \,sec^{-1}}$, we find that for {\it PIXIE}'s sensitivity the signal is below the detection limit, and even at 4 times increased sensitivity, only a marginal detection of the distortion caused by the annihilation energy release is possible. The measurements of $\Delta$ and $y_{\rm re}$ are not severely compromised by adding this possibility to the parameter estimation problem, because the additional signal is very small. 
To obtain an unambiguous $5\sigma$-detection of the p-wave annihilation signal in this scenario, the sensitivity needs to be increased $\simeq 10$ times over {\it PIXIE}.

Assuming that the relic particle is non-relativistic without any p-wave Sommerfeld enhancement one has $\left<\sigma {\rm v}\right>\propto {\rm v}^2 \propto (1+z)^2$. As mentioned above, in this case most energy is released very early causing a pure $\mu$-distortion. However, the limits from BBN and light-element abundances are expected to be much stronger, so that we do not discuss this case any further.

Next we consider energy release due to s-wave annihilation, for instance associated with a dark matter particle. The annihilation efficiency is already tightly constrained by the effect on the CMB anisotropies \citep{Peebles2000, Chen2004, Padmanabhan2005, Zhang2006}, where the best observational limit is obtained from WMAP \citep{Galli2009, Huetsi2009, Slatyer2009, Huetsi2011}, translating into $f_{\rm ann, s}\lesssim \pot{2}{-23}\,{\rm eV \, sec^{-1}}$ \citep{Chluba2010}. This case is associated with an energy release of $\Delta\rho_{\gamma}/\rho_\gamma\simeq \pot{8.3}{-9}$, available for spectral distortions. 
In contrast to the p-wave annihilation scenario, energy is liberated more evenly per logarithmic redshift interval, so that the associated spectral distortion lies between a $\mu$ and $y$-distortion (see Fig.~\ref{fig:DM_scenarios}).
Annihilations with $f_{\rm ann, s}\simeq \pot{2}{-23}\,{\rm eV \, sec^{-1}}$ remain undetectable, even at 4 times the sensitivity of {\it PIXIE}, in agreement with conclusion from previous analyses \citep{Chluba2010, Chluba2011therm}.
A $\simeq3\sigma$-detection becomes possible with 10 times the sensitivity of {\it PIXIE}.

On the other hand, assuming $f_{\rm ann, s}\simeq 10^{-22}\,{\rm eV \, sec^{-1}}$, a $\simeq 6\sigma$-detection would be possible at 4 times {\it PIXIE} sensitivity, although this scenario is already in tension with CMB anisotropy constraints.
The error for the s-wave annihilation scenario roughly scales as
\beal
\label{eq:error_s_wave}
\frac{\Delta f_{\rm ann, s}}{f_{\rm ann,s}}
& \approx 
17\%\,\left[\frac{f_{\rm ann,s}}{10^{-22}\,{\rm eV \,sec^{-1}}}\right]^{-1}\,\left[\frac{\Delta I_\nu}{4\Delta I^{\rm PIXIE}_\nu}\right]^{-1}.
\end{align}
The current limit on $f_{\rm ann, s}$ derived from CMB anisotropies may be improved by another factor of $\simeq 6$ \citep[e.g., see][for projections]{Huetsi2009, Huetsi2011} with the next release of {\it Planck} (which will include all the temperature and polarization data), {\it ACTpol} and {\it SPTpol} \citep{ACTPol, SPTpol}.
At this level of sensitivity it will be hard to directly compete using spectral distortion measurements; however, the spectral distortion constraints are independent and probe different epochs of the evolution, providing another important handle on possible systematics, e.g., related to possible uncertainties in the cosmological recombination process \citep{Farhang2011, Farhang2013}.
Additional freedom could be added due to Sommerfeld enhancement of the annihilation cross-section \citep[e.g., see][]{Hannestad2011}, but a more detailed investigation of this aspect is beyond the scope of this work.

Figure~\ref{fig:DM_scenarios} also indicates that in the p-wave annihilation scenario with $f_{\rm ann, p}\simeq 10^{-26}\,{\rm eV \, sec^{-1}}$ a similar amount of energy is deposited during hydrogen recombination ($z\simeq 10^3$) as in the well constrained s-wave annihilation scenario with $f_{\rm ann, s}\simeq \pot{2}{-23}\,{\rm eV \, sec^{-1}}$. 
We thus did not consider cases with larger p-wave annihilation cross-section, because these would already be in tension with the CMB anisotropy data. Improving the limit on p-wave annihilation scenarios with CMB anisotropy measurements will, however, be very hard and the distortion signal has a larger leverage, offering a way to detect the signatures from particles with p-wave annihilation efficiency $f_{\rm ann, p}\gtrsim \pot{\rm few}{-28}\,{\rm eV \, sec^{-1}}$ at {\it PIXIE}'s sensitivity.

\begin{figure}
\centering
\setlength{\unitlength}{0.05\columnwidth}
  \begin{picture}(20,20)
    \put(0,0){\includegraphics[width=\columnwidth]{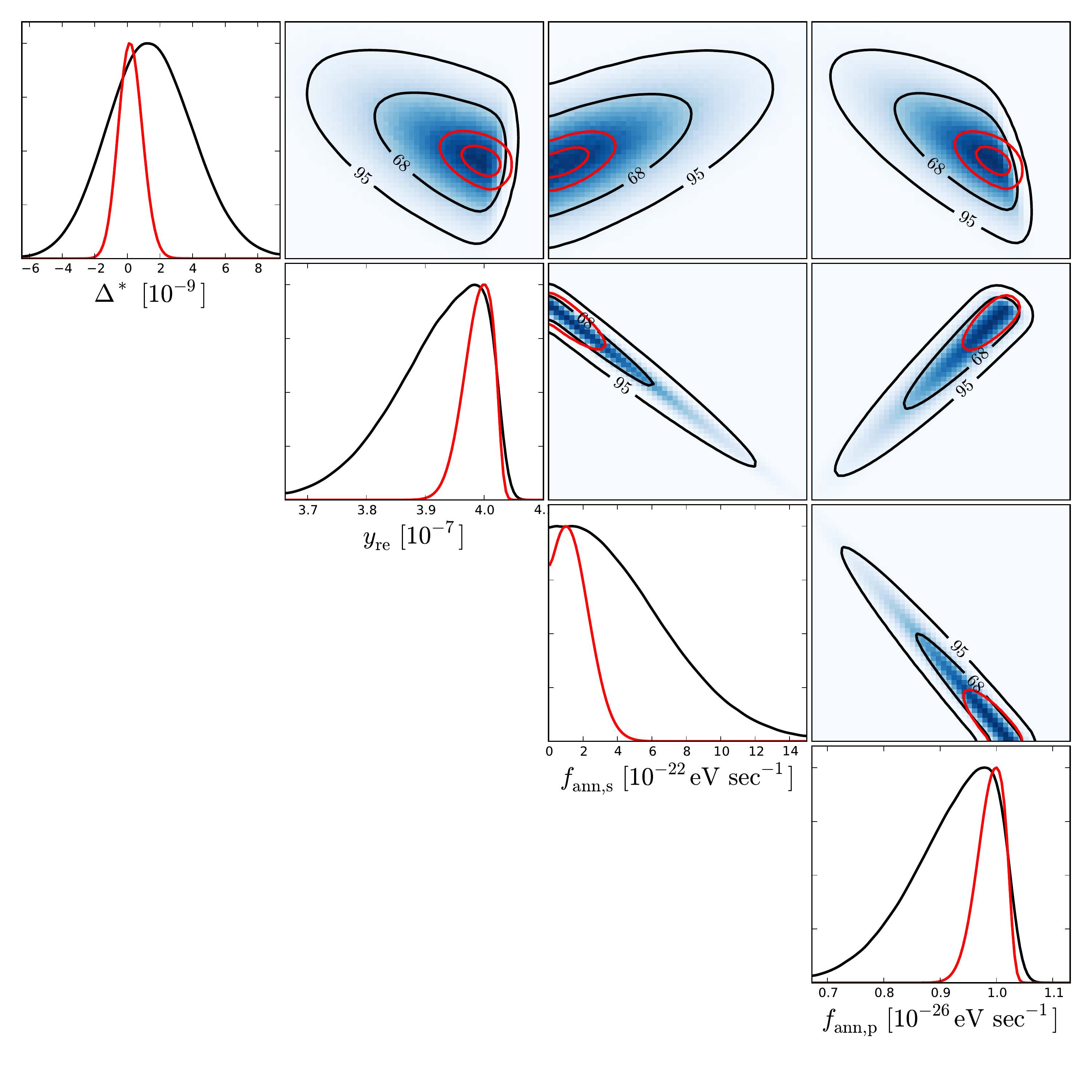}}
    \put(1.0,13.5){$(\Delta^\ast\equiv\Delta-\Delta_{\rm f})$}
    \put(1.1,7.3){Fiducial values:}
    \put(1.1,6.1){$\Delta_{\rm f}=\pot{1.2}{-4}$}
    \put(1.1,4.9){$y_{\rm re}=\pot{4}{-7}$}
    \put(1.1,3.7){$f_{\rm ann, s}=10^{-22}\,{\rm eV \, sec^{-1}}$} 
    \put(1.1,2.5){$f_{\rm ann, p}=10^{-26}\,{\rm eV \, sec^{-1}}$}
  \end{picture}
%
\caption{Large distortion s- and p-wave annihilation scenario. Contours and lines are as before. Degeneracies between the parameters prevent a distinction of the signatures of both particles, even for high sensitivity.}
\label{fig:s_p_mixed}
\end{figure}
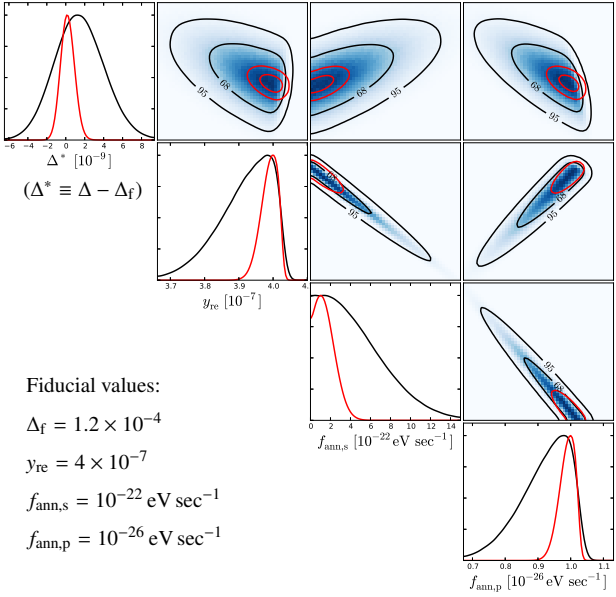

Finally, in Fig.~\ref{fig:s_p_mixed} for illustration we show the large distortion scenario ($f_{\rm ann, s}\simeq 10^{-22}\,{\rm eV \, sec^{-1}}$ and $f_{\rm ann, p}\simeq 10^{-26}\,{\rm eV \, sec^{-1}}$) of Fig.~\ref{fig:DM_scenarios}, with simultaneous energy release due to particles with s- and p-wave annihilation. The parameters becomes rather degenerate, and a separate detection of the s-wave annihilation effect remains challenging even at 4 times the sensitivity of {\it PIXIE}. Although an individual detection of the s- or p-wave annihilation signature should be possible, the two signals are simply too similar and strong correlations cause large uncertainties and biases in the parameters, which only disappear at high sensitivity. This makes the projected 2D probability distributions shown in Fig.~\ref{fig:s_p_mixed} very non-Gaussian. At $\simeq 20$ times the sensitivity of {\it PIXIE}, we find a $\simeq 2\sigma$ detection of the s-wave annihilation signature and $f_{\rm ann, p}\simeq 1\%$ from the p-wave annihilation signal.

Considering a small distortion scenario with more comparable contributions from s- and p-wave annihilations ($f_{\rm ann, s}\simeq \pot{2}{-23}\,{\rm eV \, sec^{-1}}$ and $f_{\rm ann, p}\simeq 10^{-28}\,{\rm eV \, sec^{-1}}$), we find that an improvement of the sensitivity by a factor of $\simeq 40$ is needed to start distinguishing the signals from both particles, rendering an analysis along these lines more futuristic.
This is because for this scenario the signal is close to the detection limit of {\it PIXIE}, and the differences with respect to a pure superposition of $\mu$- and $y$-distortions, which could be used to distinguish the two cases, are only a small correction, necessitating this large improvement of the sensitivity.

\begin{figure}
\centering
\setlength{\unitlength}{0.05\columnwidth}
  \begin{picture}(20,20)
    \put(0,0){\includegraphics[width=\columnwidth]{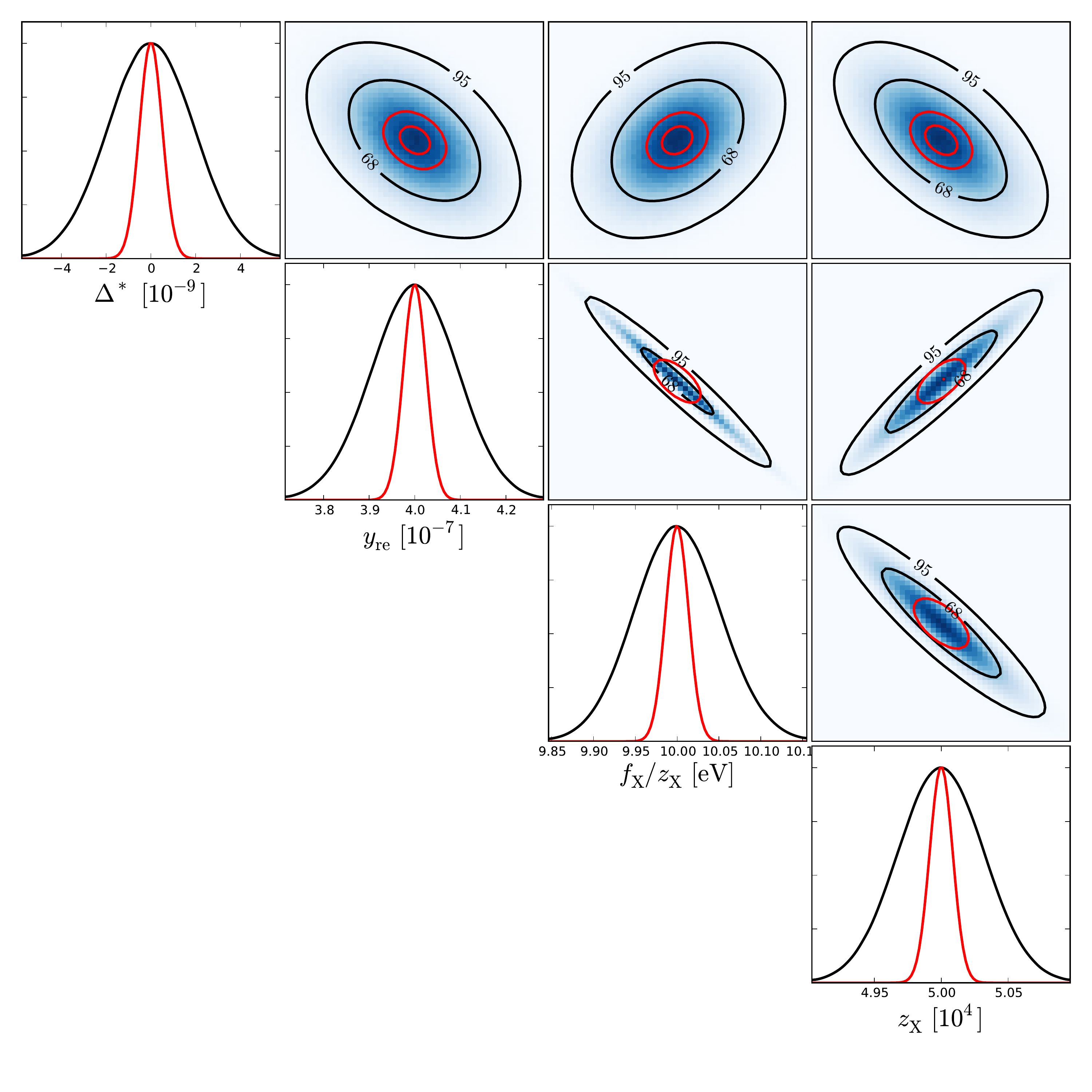}}
    \put(1.0,13.5){$(\Delta^\ast\equiv\Delta-\Delta_{\rm f})$}
    \put(1.1,7.3){Fiducial values:}
    \put(1.1,6.1){$\Delta_{\rm f}=\pot{1.2}{-4}$}
    \put(1.1,4.9){$y_{\rm re}=\pot{4}{-7}$}
    \put(1.1,3.7){$f_{\rm X}=\pot{5}{5}\,{\rm eV}$} 
    \put(1.1,2.5){$z_{\rm X}=\pot{5}{4}$ ($\Gamma_{\rm X}\simeq \pot{1.1}{-8}{\rm sec^{-1}}$)}
  \end{picture}
\\
  \begin{picture}(20,20)
    \put(0,0){\includegraphics[width=\columnwidth]{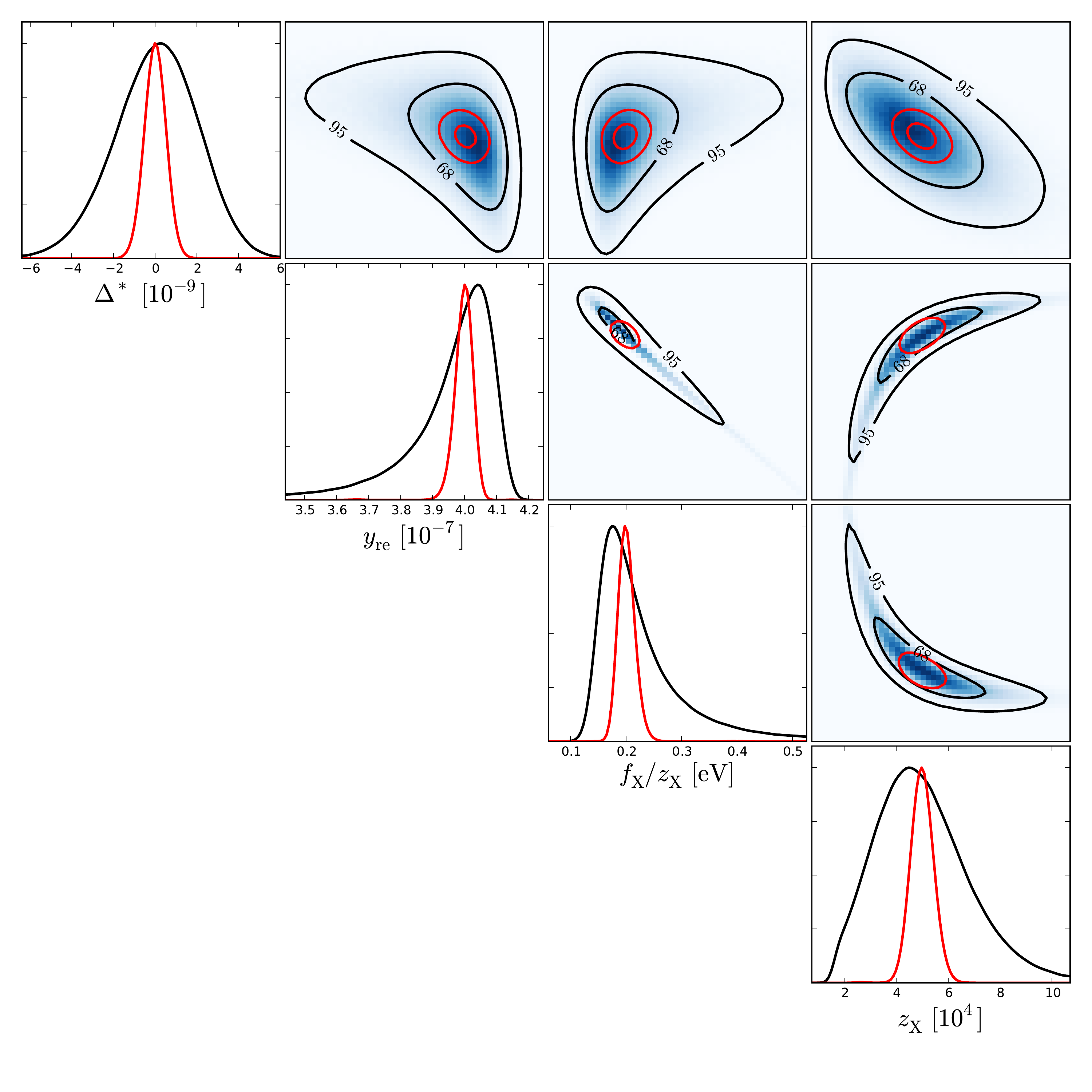}}
    \put(1.0,13.5){$(\Delta^\ast\equiv\Delta-\Delta_{\rm f})$}
    \put(1.1,7.3){Fiducial values:}
    \put(1.1,6.1){$\Delta_{\rm f}=\pot{1.2}{-4}$}
    \put(1.1,4.9){$y_{\rm re}=\pot{4}{-7}$}
    \put(1.1,3.7){$f_{\rm X}=10^{4}\,{\rm eV}$} 
    \put(1.1,2.5){$z_{\rm X}=\pot{5}{4}$ ($\Gamma_{\rm X}\simeq \pot{1.1}{-8}{\rm sec^{-1}}$)}
  \end{picture}
\caption{Large- and small-distortion decaying particle scenario. Contours and lines are as before. For large energy release the distortion can be easily constrained; however, for small energy release the parameter space becomes more complicated and higher sensitivity improves matters significantly.}
\label{fig:Decay}
\end{figure}

\section{Decaying particle scenarios}
\label{sec:decay}
Decaying relic particles with lifetimes $\simeq 380\,{\rm kyr}$ (corresponding to the time of recombination) are again tightly constrained by measurement of the CMB anisotropies \citep{Zhang2007, Giesen2012}, while particles with lifetimes comparable to minutes can affect the light-element abundances and bounds derived from BBN apply \citep{Kawasaki2005, Jedamzik2008}.
However, experimental constraints for particles with lifetimes $\simeq 10^{6}-10^{12}\, {\rm sec}$ are less stringent, still leaving rather large room for extra energy release $\Delta\rho_{\gamma}/\rho_\gamma\lesssim 10^{-6}-10^{-5}$ \citep[e.g.,][]{Hu1993b, Kogut2011PIXIE}. Large energy-release rates are especially possible for very light particles with masses $\lesssim {\rm MeV}$.
A {\it PIXIE}-type CMB experiment thus has a large potential to discover the signature of some long-lived relic particles or at least provide complementary and independent constraints to these scenarios.
If most of the energy is released at $z\gtrsim \pot{3}{5}$, a pure $\mu$-distortion is created, so that this case is practically degenerate, e.g., with scenarios that include an annihilating particle with p-wave annihilation cross-section. However, for energy release around $z\simeq \pot{5}{4}$, the distortion can differ sufficiently to become distinguishable.

In Fig.~\ref{fig:Decay}, we show the projected constraints for a large- and small-distortion scenario, with energy release $\Delta\rho_{\gamma}/\rho_\gamma\simeq \pot{6.4}{-6}$ and $\Delta\rho_{\gamma}/\rho_\gamma\simeq \pot{1.3}{-7}$, respectively. Since the total energy release scales as $\Delta\rho_\gamma/\rho_\gamma\propto f_{\rm X}/z_{\rm X}$ \citep[cf.][]{Chluba2011therm}, it is best to consider the variables $f_{\rm X}/z_{\rm X}$ and $z_{\rm X}\simeq \pot{4.8}{9}\,\Gamma_{\rm X}^{1/2}\,{\rm sec}^{1/2}$ as parameters. This reduces the parameter covariance significantly. To accelerate the computation, we furthermore tabulate the distortion for different particle lifetimes and interpolate on this grid to obtain the resulting distortion. 
With this method $\simeq \pot{5}{6}$ samples can be taken in a few minutes on a standard quad-core laptop.

\begin{figure}
\centering
\hspace{-3mm}\includegraphics[width=1.02\columnwidth]{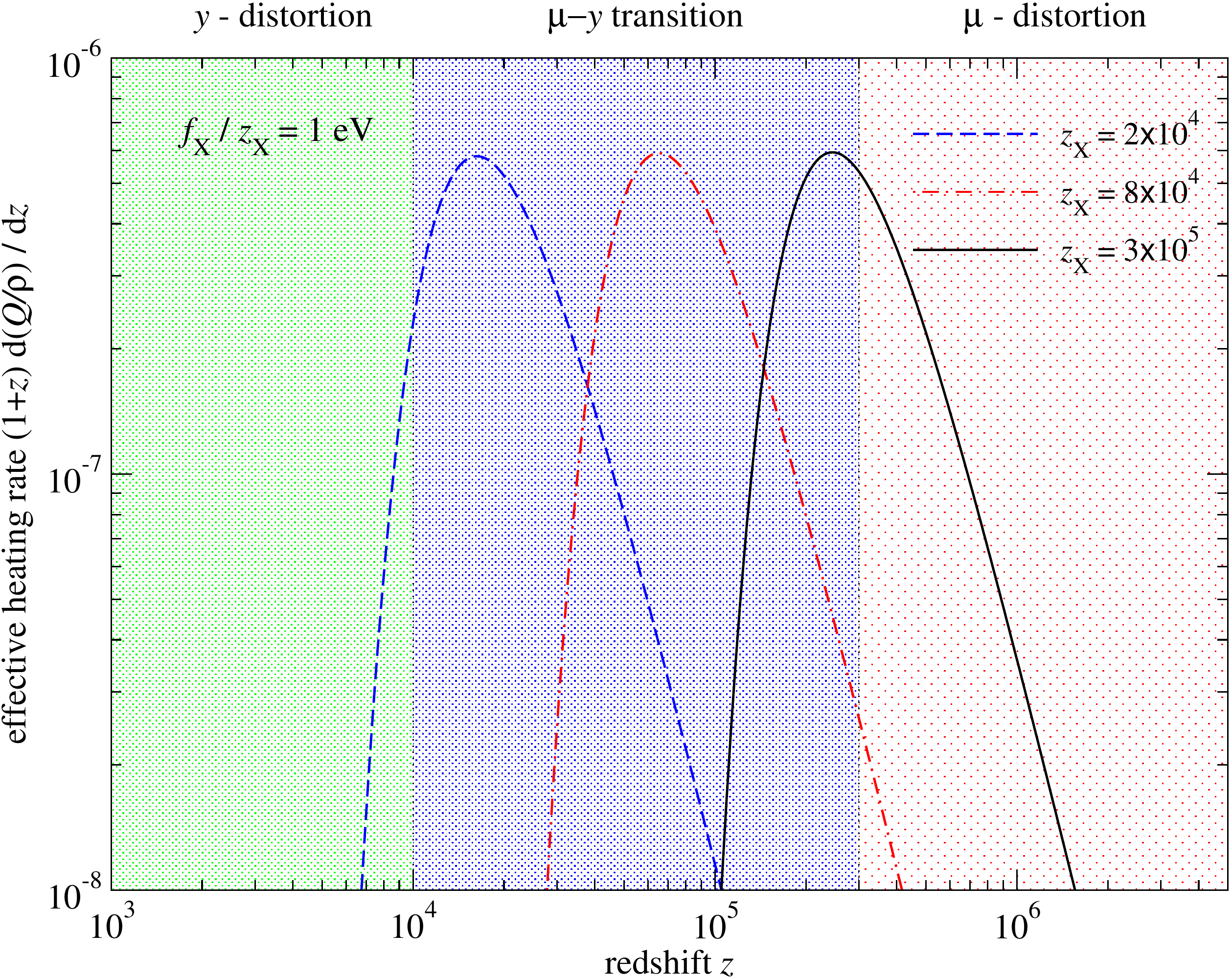}
\\[4mm]
\hspace{-3mm}\includegraphics[width=1.02\columnwidth]{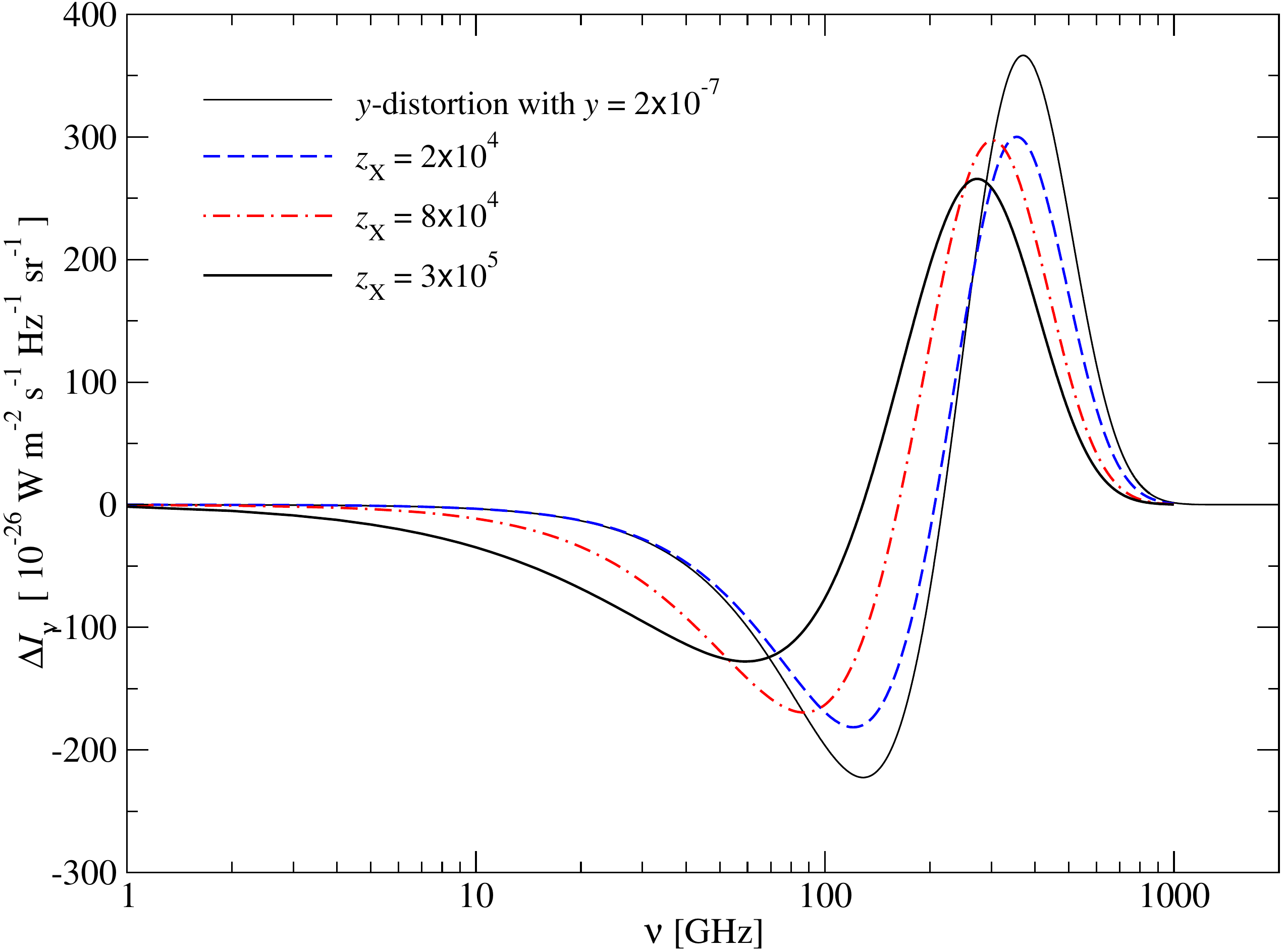}
\\[4mm]
\hspace{-3mm}\includegraphics[width=1.02\columnwidth]{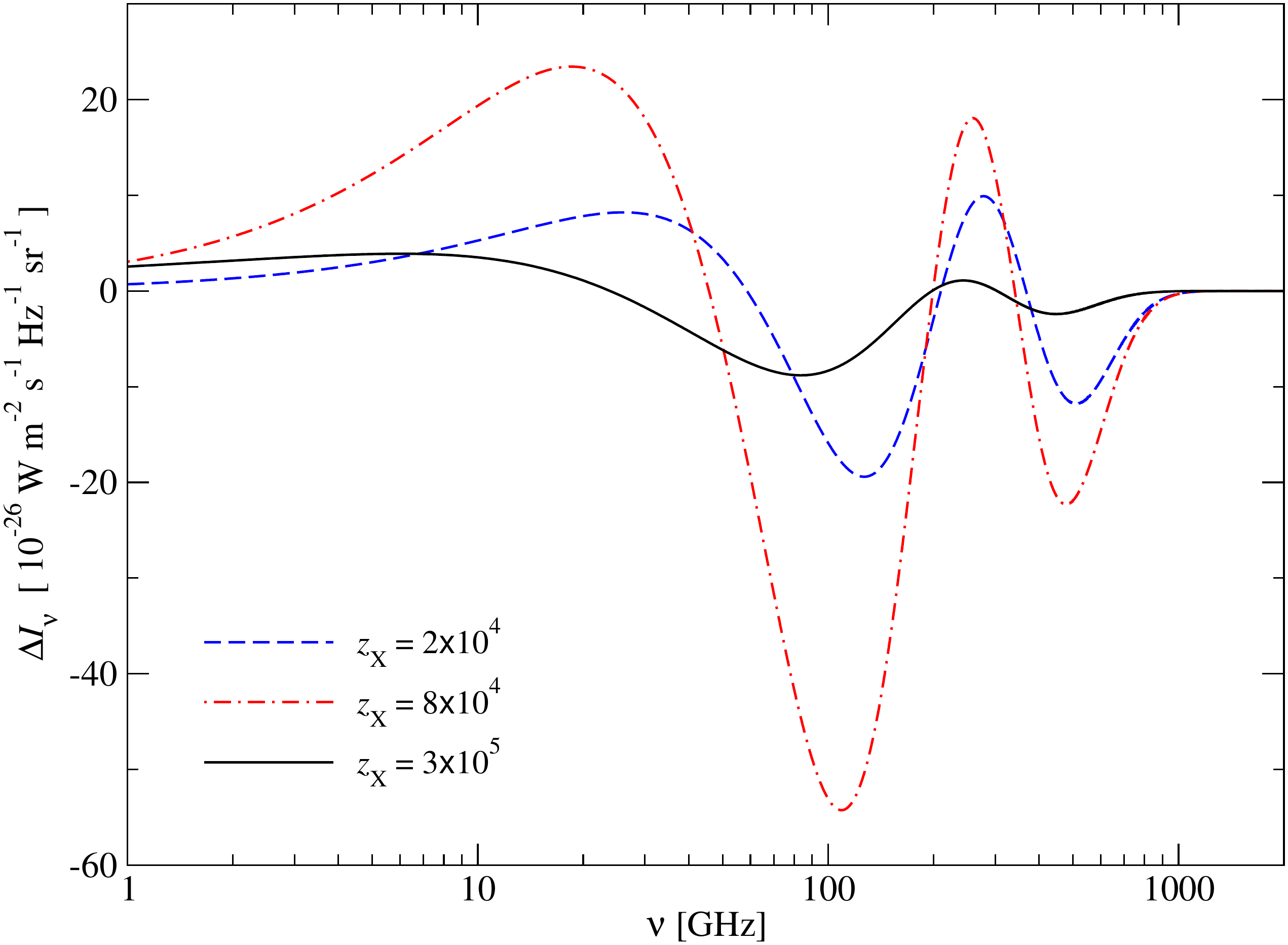}
\caption{Lifetime effect for different decaying particle scenarios. The upper panel shows the energy-release rate for all cases, while the central panel illustrates the distortion in comparison with a $y$-distortion of $y=\pot{2}{-7}$. The lower panel shows the residual distortion after subtracting the best-fitting $\mu$- and $y$-distortions.}
\label{fig:Decay_scenarios}
\end{figure}

From Fig.~\ref{fig:Decay}, one can see that for the large-distortion scenario, a $\simeq 1\%$ precision can be achieved for $f_{\rm X}/z_{\rm X}$ and $z_{\rm X}$ assuming {\it PIXIE} sensitivity. The uncertainty of $y_{\rm re}$ increases from $\simeq 1\%$ in the p-wave scenario (see Fig.~\ref{fig:pwave_DM}) to about $\simeq 3\%$ due to correlations with the signal induced by the decaying particle. This is simply because a noticeable fraction of the decay energy goes into production of $y$-distortions at late time, which induces an anti-correlation of $f_{\rm X}/z_{\rm X}$ and $y_{\rm re}$, but a correlation of $z_{\rm X}$ and $y_{\rm re}$ (increasing $z_{\rm X}$ means less energy release at low redshift close to recombination, and hence more of the $y$-distortion is attributed to $y_{\rm re}$).

Considering the small-distortion scenario (with $\Delta\rho_{\gamma}/\rho_\gamma\simeq \pot{1.3}{-7}$ going into distortions) shows that at {\it PIXIE} sensitivity the parameter space becomes rather large, showing extended regions of low probability due to degeneracies and correlations. Improving the sensitivity four times significantly tightens possible constraints on these scenarios, allowing better than $5\sigma$-detections of the particle signature.
The constraint on the amount of energy that is released ($\propto f_{\rm X}/z_{\rm X}$) is less prone to changes in the sensitivity than $z_{\rm X}$, being a proxy for the particle's lifetime. This is because sensitivity to $z_{\rm X}$ is introduced mainly by the ability to distinguish a superposition of pure $\mu$- and $y$-distortion from full distortion \citep{Chluba2011therm, Khatri2012mix, Chluba2013Green}, but the residuals are a correction and thus harder to utilize.

One can ask the question about how strongly the errors change when leaving the total energy release constant, but varying the particle lifetime. For $z_{\rm X}\lesssim 10^4$, one expects degeneracy with the $y$-distortion created at low redshifts, while for lifetimes shorter than $t_{\rm X}\simeq \pot{3}{8}\,{\rm sec}$ the signal becomes maximally orthogonal ($\mu$- versus $y$-distortion). 
In Fig.~\ref{fig:Decay_scenarios}, we illustrate this dependence of the spectral distortion on $z_{\rm X}$. Decreasing $z_{\rm X}$ (i.e., increasing the lifetime) moves the distortion from $\mu$- to a $y$-distortion. The residual of the distortion with respect to a superposition of pure $\mu$- and $y$-distortions is largest for $z_{\rm X}\simeq\pot{8}{4}$, reaching roughly $30\%$ of the total signal at $100\,{\rm GHz}$. Both closer to $z_{\rm X}\simeq 10^4$ and $z_{\rm X}\simeq \pot{3}{5}$, the residual becomes smaller, making a distinction harder.

Assuming $f_{\rm X}/z_{\rm X}=1\,{\rm eV}$ and {\it PIXIE} sensitivity, we find that for $z_{\rm X}\simeq \pot{4}{4}-\pot{2}{5}$ the errors on $f_{\rm X}/z_{\rm X}$ and $z_{\rm X}$ are typically better than $\lesssim 30\%$. At $z_{\rm X}\simeq 10^5$, we obtain a $\simeq 8\%$ error on $f_{\rm X}/z_{\rm X}$ and $\simeq 6\%$ error on $z_{\rm X}$, representing one of the best cases. 
For $z_{\rm X}\lesssim \pot{4}{4}$, the degeneracy with $y_{\rm re}$ already becomes too large and the error on $z_{\rm X}$ inflates to $\gtrsim 27\%$. Similarly, for $z_{\rm X}\gtrsim \pot{2}{5}$, the signal is already too close to a pure $\mu$-distortion, which causes a large degeneracy between $f_{\rm X}/z_{\rm X}$ and $z_{\rm X}$ (with multimodal solutions), simply because simultaneously increasing $f_{\rm X}/z_{\rm X}$ and $z_{\rm X}$ (to compensate for the suppression of the distortion amplitude by thermalization) gives rise to the same distortion. In other words, a pure $\mu$-distortion is insensitive to when it was created and thus does not allow differentiating between scenarios with different particle lifetimes at $z\gtrsim \pot{\rm few}{5}$.
Still, a tight upper limit on the total amount of energy that is released can be placed, constraining the possible abundance of decaying particles with lifetimes $t_{\rm X}\simeq \pot{6}{6}\,{\rm sec}-\pot{3}{8}\,{\rm sec}$.

These statements, however, depend strongly on the sensitivity of the experiment and on how large the average distortion is. 
As explained above, the information about the particle lifetime is largely encoded in the deviations from a pure superposition of $\mu$- and $y$-distortions; however, the residual is a correction and thus higher sensitivity or a larger distortion are needed to make use of that information.
Assuming $f_{\rm X}/z_{\rm X}=1\,{\rm eV}$ and $z_{\rm X}=\pot{2}{4}$, a {\it PIXIE}-type experiment is unable to constrain the lifetime of the particle. The degeneracy is already broken at twice the sensitivity of {\it PIXIE}, yielding $\simeq 29\%$ error on $f_{\rm X}/z_{\rm X}$ and $\simeq 17\%$ error on $z_{\rm X}$. This further improves to  $\simeq 14\%$ uncertainty in $f_{\rm X}/z_{\rm X}$ and a $\simeq 9\%$ error on $z_{\rm X}$ for four times the sensitivity of {\it PIXIE}.
This energy-release scenario corresponds to $\Delta\rho_{\gamma}/\rho_\gamma\simeq \pot{6.4}{-7}$, such that the distortion is comparable in amplitude to the $y$-signal from late times. Assuming that less energy is liberated by the decaying particle increases the errors (and hence the degeneracy), and conversely, for larger decay energy the errors diminish. 
Overall, a {\it PIXIE}-type experiment will provide a pretty good probe for long-lived particles with lifetimes $t_{\rm X}\simeq \pot{6}{8}\,{\rm sec}-10^{10}\,{\rm sec}$ and $f_{\rm X}/z_{\rm X}\gtrsim 1\,{\rm eV}$.

\section{Dissipation of small-scale acoustic modes}
The prospect of accurate measurements of the CMB spectrum with a {\it PIXIE}-type experiment spurred renewed interests in how primordial perturbations at small scales dissipate their energy \citep{Chluba2011therm, Khatri2011BE, Pajer2012, Chluba2012, Dent2012, Ganc2012, Chluba2012inflaton, Powell2012, Khatri2013forecast, Chluba2013iso}.
It was shown, that this effect can be used to place tight limits on the amplitude and shape of the power spectrum at scales far smaller than what is probed with measurements of the CMB anisotropies, in principle allowing us to discover the distortion signatures from several classes of early-universe models \citep[e.g., see][]{Chluba2012inflaton}.

\begin{figure}
\centering
\includegraphics[width=\columnwidth]{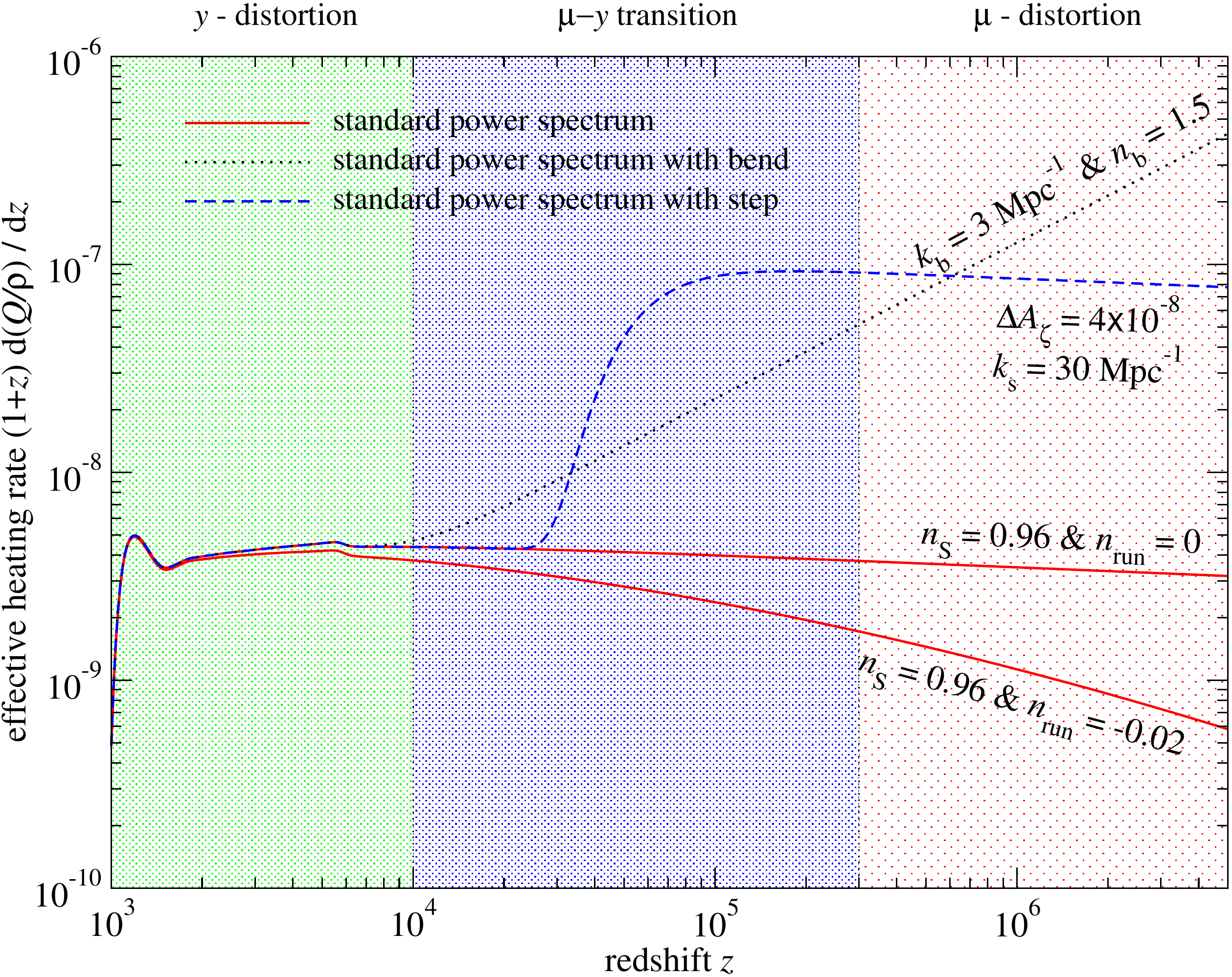}
\\[4mm]
\includegraphics[width=\columnwidth]{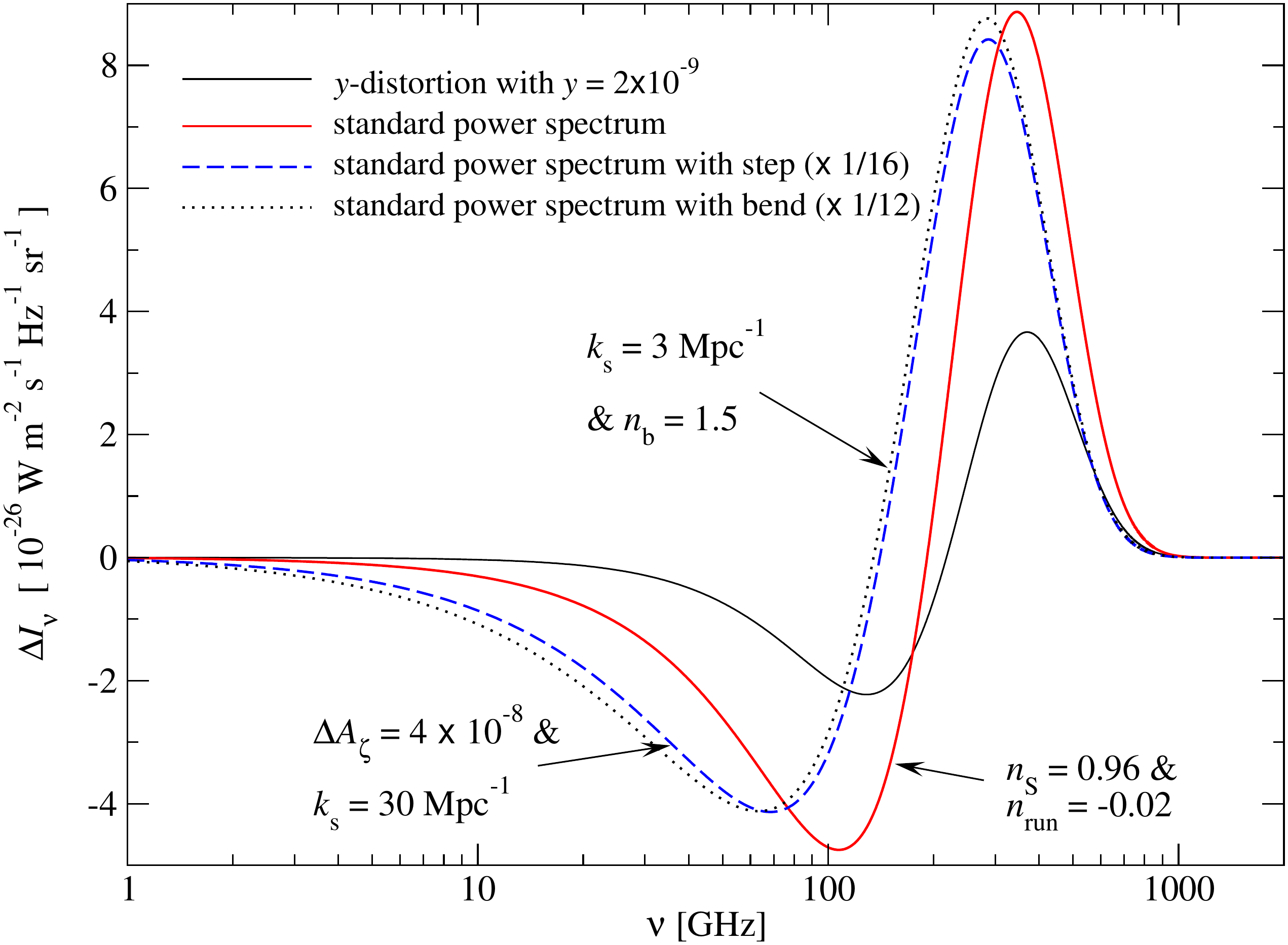}
\caption{Effective heating rate (upper panel) and associated spectral distortion (lower panel) caused by the dissipation of small-scale acoustic modes in different scenarios. For reference, we show a $y$-distortion with $y=\pot{2}{-9}$. For the standard power spectrum, we used $A_\zeta=\pot{2.2}{-9}$ and $\nS=0.96$ at pivot scale $k_0=0.05\,\Mpc^{-1}$. All but one case are without running. The two scenarios with a step and bend of the primordial power spectrum lead to rather similar distortions (modulo and overall factor), and thus become hard to distinguish, although each model should be detectable with a {\it PIXIE}-like experiment at more than $5\sigma$-confidence.}
\label{fig:diss_scenarios}
\end{figure}

Taking a conservative perspective, one can assume that the power spectrum of curvature perturbations is fully determined by CMB anisotropy measurements at large scales, implying an amplitude $A_\zeta\simeq \pot{2.2}{-9}$, spectral index $\nS\simeq 0.96$ and its running $\nrun\simeq -0.02$, at pivot scale $k_0=0.05\,\Mpc^{-1}$ \citep{Planck2013params}.
This is a significant extrapolation from wavenumbers $k<1\,\Mpc^{-1}$ all the way to $k\simeq \pot{\rm few}{4}\,\Mpc^{-1}$, and it was already argued that for a {\it PIXIE}-type experiment the signal remains just short of the $1\sigma$-detection limit \citep{Chluba2011therm, Chluba2012}. 
Improving the sensitivity a few times will allow a detection of this signal; however, given that the errors on $A_\zeta$ and $\nS$ from CMB data are now $\lesssim 1\%$, to use spectral distortion {\it alone} as a competitive probe, we find that a factor of $\simeq 100-200$ improvement in the sensitivity is necessary. The strongest dependence of the distortion signal is due to $\nrun$ (see Fig.~\ref{fig:diss_scenarios} for illustration), since small changes affect the amplitude of the small-scale power spectrum and hence the associated spectral distortion by a large amount \citep{Khatri2011BE, Chluba2012}, providing some amplification and the possibility to break parameter degeneracies. Still, this application of spectral distortion measurements remains more futuristic, being comparable to the challenge of measuring the cosmological hydrogen and helium recombination features with high precision.

Both from the theoretical and observational point of view, there is, however, no reason to believe that the small-scale power spectrum is described by what is dictated by large-scale measurements. There is no shortage of models that create, bumps, kinks, steps, or oscillatory features in the primordial power spectrum \citep[e.g.,][]{1989PhRvD..40.1753S, 1992JETPL..55..489S, 1994PhRvD..50.7173I, 1996NuPhB.472..377R, Stewart1997b, 1998PhRvD..58f3508C,1998GrCo....4S..88S, 2000PhRvD..62d3508C, Hunt:2007dn, 2008PhRvD..77b3514J, Neil2009I, Barnaby2010, Ido2010, 2011JCAP...01..030A,2012arXiv1201.4848C}, and direct observational constraints \citep[e.g., see][for overview]{BSA11} leave large room for excess power at $k\gtrsim {\rm few}\times \Mpc^{-1}$.
The recent results obtained with {\it Planck}, e.g., from limits to non-Gaussianity \citep{Planck2013ng}, certainly further reduce the allowed parameter space for different models, but the existence of large-scale anomalies \citep{Planck2013power} and possible small-scale power spectrum features \citep{Planck2013iso} indicate that matters might be more complex.
A {\it PIXIE}-type experiment will therefore open up a new window to early-universe models, no matter if a distortion is detected or not.

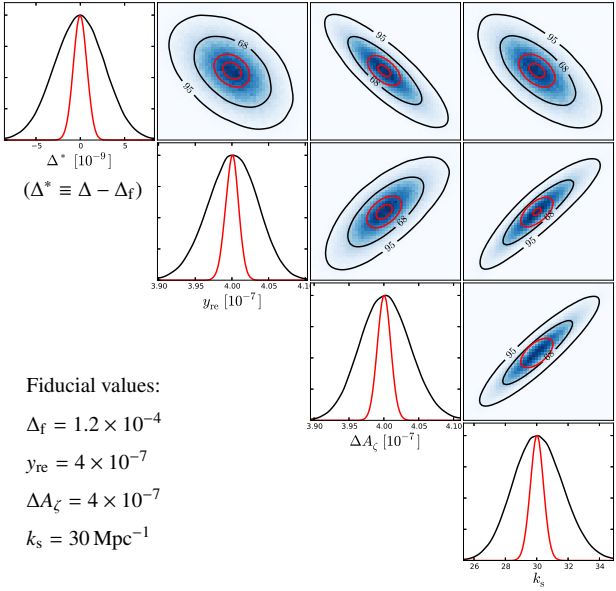
\begin{figure}
\centering
\setlength{\unitlength}{0.05\columnwidth}
  \begin{picture}(20,20)
    \put(0,0){\includegraphics[width=\columnwidth]{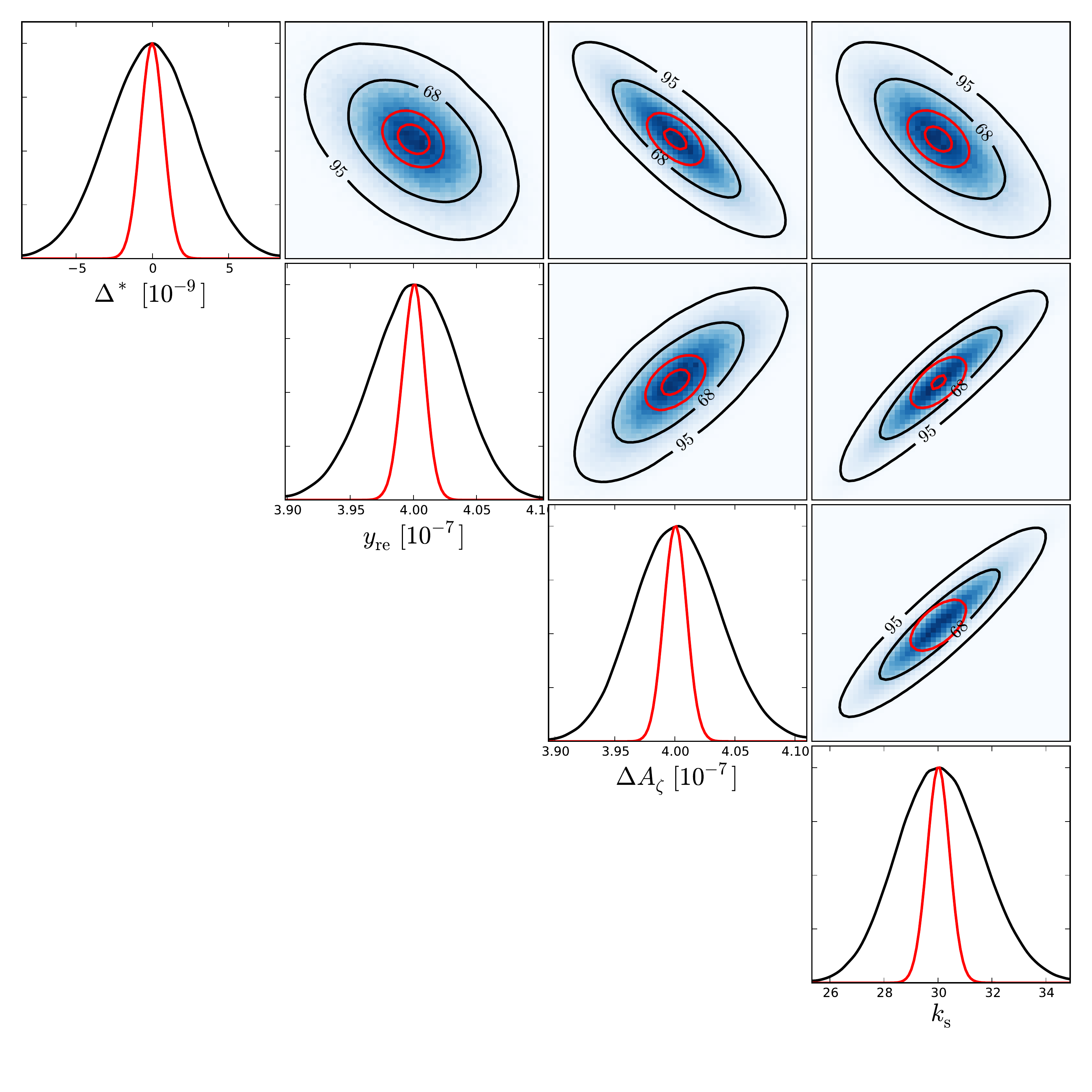}}
    \put(1.0,13.5){$(\Delta^\ast\equiv\Delta-\Delta_{\rm f})$}
    \put(1.1,7.3){Fiducial values:}
    \put(1.1,6.1){$\Delta_{\rm f}=\pot{1.2}{-4}$}
    \put(1.1,4.9){$y_{\rm re}=\pot{4}{-7}$}
    \put(1.1,3.7){$\Delta A_\zeta=\pot{4}{-7}$} 
    \put(1.1,2.5){$k_{\rm s}=30\,\Mpc^{-1}$}
  \end{picture}
\caption{Large-distortion scenario caused by a step in the small-scale power spectrum. Contours and lines are as before. A {\it PIXIE}-like experiment allows constraining scale and amplitude of a step in the power spectrum at $k_{\rm s}\simeq 20\,\Mpc^{-1}-50\,\Mpc^{-1}$ with $\Delta A_\zeta\gtrsim \pot{4}{-7}$ to $\lesssim 6\%$ precision.}
\label{fig:diss_constraint}
\end{figure}
Given the range of possibilities, we shall pick a few illustrative cases, representing simple classes of models. Detailed constraints on specific models should be derived in a case-by-case basis; however, our selection provides some intuition for what could be possible in the future. 
We start with a simple step, $\Delta A_\zeta>0$, in the amplitude of the curvature power spectrum at different $k\gtrsim {\rm few}\times \Mpc^{-1}$, assuming a spectral index $\nS'$. If $\nS'\simeq 1$ and $k_{\rm s}\simeq 3\,\Mpc^{-1}$, from the practical point of view this case is degenerate with the spectral distortion produced by s-wave annihilation [both have a heating rate $\id (Q/\rho_\gamma)/\id z\propto z^{-1}$ over most redshifts]. The difference is, however, that a step in the small-scale power spectrum at $k_{\rm s}\simeq  3\,\Mpc^{-1}$ is not as tightly constrained by large-scale CMB anisotropy measurements, but could be tightly constrained with a {\it PIXIE}-type experiment \citep{Chluba2012inflaton}.
More generally, degeneracy with annihilation scenarios and $\left<\sigma {\rm v}\right>\propto (1+z)^k$ exists if $\nS'\simeq k+1$. For simplicity, below we consider only the case $\nS'\equiv \nS$.

In Fig.~\ref{fig:diss_scenarios}, we show the heating rate and expected distortion for this scenario, assuming $\Delta A_\zeta\simeq \pot{4}{-8}$, $k_{\rm s}=30\,\Mpc^{-1}$ and $\nS'=0.96$ for illustration. At $z\gtrsim \pot{5}{5}$, the effective heating rate is $\simeq 25$ times larger than that for the standard background power spectrum without running. Consequently, also the $\mu$-type contribution to the resulting spectral distortion is found to be $\simeq 16$ times larger, with additional contributions from the $\mu-y$-transition era. 
We can also see that the effective heating rate changes gradually to the one of the background model around $z\simeq \pot{4}{4}$. Modes with fixed wavenumber $k$ dissipate their energy in a range of redshifts with a maximum at redshifts $z_{\rm diss}\simeq \pot{4.5}{5}[k/10^3\,\Mpc^{-1}]^{2/3}$ \citep{Chluba2012inflaton}. Thus, no abrupt change of the heating rate is expected.

Since the distortion in principle depends on how energy is released at $10^4 \lesssim z\lesssim \pot{3}{5}$, one does expect to be sensitive to $k_{\rm s}$. 
From the discussion of decaying particle scenarios, it is already clear that only for rather large distortions (i.e., a step amplitude $\Delta A_\zeta\gtrsim \pot{\rm few}{-7}$) will a {\it PIXIE}-type experiment be able to constrain the position of the step.
In Fig.~\ref{fig:diss_constraint}, we show the projected constraints on this scenario, assuming that $\nS'=0.96$ is fixed with $\Delta A_\zeta =  \pot{4}{-7}$ and $k_{\rm s}=30\,\Mpc^{-1}$. Both the amplitude and position of the step are well constrained, with $\Delta k_{\rm s}/k_{\rm s}\simeq 5\%$ and $\Delta\Delta A_\zeta/\Delta A_\zeta\simeq 1\%$.
Increasing the sensitivity (or similarly considering a scenario with larger step amplitude) further tightens the constraints. 
Similar to the discussion for the decaying particle case, moving $k_{\rm s}$ closer to $\simeq 3\,\Mpc^{-1}$ ($z_{\rm diss}\simeq 10^4$) or $\simeq 540\,\Mpc^{-1}$ ($z_{\rm diss}\simeq \pot{3}{5}$), the sensitivity on the position of the step is expected to degrade. A {\it PIXIE}-type experiment will be most sensitive to a step at $k_{\rm s}\simeq 20\,\Mpc^{-1}-50\,\Mpc^{-1}$ giving $\Delta k_{\rm s}/k_{\rm s}\lesssim 6\%$ and $\Delta\Delta A_\zeta/\Delta A_\zeta\lesssim 2\%$ for $\Delta A_\zeta\gtrsim \pot{4}{-7}$.
For $k_{\rm s}\gtrsim 150\,\Mpc^{-1}$ and $k_{\rm s}\lesssim 2\,\Mpc^{-1}$, the error in $k_{\rm s}$ increases above $\simeq 30\%$, although at the boundaries, the amplitude of the step can still be constrained rather precisely ($\simeq 6\%$ and $\simeq 1\%$, respectively). This is because the distortion is rather large and only the information about the position of the step is lost at these limits.
Improved sensitivity again helps breaking degeneracies, allowing us to tighten the constraints on this scenario and broadening the range over which the location of the step can be determined, analogous to the decaying particle case.

As an additional example, we shall consider a primordial power spectrum with a change of the spectral index from $\nS$ to $n_{\rm b}$ at some scale $k_{\rm b}$, introducing a simple bend. This behavior could be expected from running mass models \citep[e.g.,][]{Stewart1997, Stewart1997b, 1999PhRvD..59f3515C} or a small-scale isocurvature mode with blue spectral index that is completely subdominant at large scales \citep{Chluba2013iso}.
The shape of the distortion is determined by $n_{\rm b}$, since it sets the $\mu/y$-distortion mixture and how significant deviations from this simple superposition are.
The value of $k_{\rm b}$ just parametrizes the overall amplitude of the distortion, with spectral distortion measurements being insensitive to scenarios with $k_{\rm b}\gtrsim \pot{\rm few}{4}\,\Mpc^{-1}$ \citep{Chluba2012inflaton}.
Also, if $n_{\rm b}<\nS$, this scenario will be hard to constrain, since already the signature from the standard power spectrum is rather small. For $n_{\rm b}>\nS$, more energy is dissipated and hence a detection should be possible with a {\it PIXIE}-like experiment (see Fig.~\ref{fig:diss_scenarios} for illustration).

Estimates for the amplitude of the distortion can be computed from the model of the small-scale power spectrum, using Eq.~\eqref{eq:Q_ac_eff} and $\Delta \rho_\gamma/\rho_\gamma\simeq \int \id (Q/\rho_\gamma)/\id z \, \expf{-(z/\zmu)^{5/2}} \id z$.
Furthermore, an approximation of how large the effective $\mu$- and $y$-parameters are can be obtained using the $k$-space window function given in \citet{Chluba2012inflaton} and \citet{Chluba2013iso} or the simple approximations for the Green's function provided in \citet{Chluba2013Green}.
From these considerations, it follows that the larger $k_{\rm b}$, the steeper does the small-scale power spectrum have to become for fixed experimental sensitivity to allow determination of the spectral index and bend location. Similarly, for a fixed value of $n_{\rm b}$, sensitivity to the location of the bend is diminished, the larger $k_{\rm b}$ becomes (the distortion becomes smaller since less energy is liberated).

For parameter estimations, it is better to specify the amplitude of the small-scale power spectrum at some pivot scale instead of using $k_{\rm b}$. Setting the power spectrum amplitude, $A_{\rm b}$, at $k_{0,\rm b}\simeq 45\,\Mpc^{-1}$ keeps the total energy release roughly constant when changing $n_{\rm b}$. 
%
%
To determine $A_{\rm b}(k_{\rm b})$, assuming no running of the background power spectrum, we can use $A_{\rm b}=A_\zeta k_{\rm b}^{\nS-n_{\rm b}}\,k_0^{1-\nS}\,k_{0,\rm b}^{n_{\rm b}-1}$.
To ensure that $k_{\rm b}\geq 1\,\Mpc^{-1}$ (we shall not consider cases with both step and change of the spectral index here), we have the condition $A_{\rm b}\lesssim \pot{2.0}{-9}\,45^{n_{\rm b}-1}$.

To give some examples, for $n_{\rm b}=1.5$ and $k_{\rm b}= 3\,\Mpc^{-1} (A_{\rm b}\simeq \pot{7.2}{-9})$, the total energy  available for creation of distortions is $\Delta \rho_\gamma/\rho_\gamma\simeq \pot{2.5}{-7}$. The associated signal should be easily detectable with a {\it PIXIE}-like experiment; however, due to degeneracies the underlying parameters are less constrained. We find that for 10 times {\it PIXIE}'s sensitivity, the errors are $\Delta n_{\rm b}/n_{\rm b}\simeq 10\%$ and $\Delta A_{\rm b}/A_{\rm b}\simeq 15\%$ around the most probable solution; however, additional low-probability solutions away from the input parameters were found, showing how challenging it is to constrain this scenario even at this sensitivity.

Matters do not improve when assuming a larger change of the spectral index. For instance, for a scenario with $n_{\rm b}=2$ and $k_{\rm b}= 3\,\Mpc^{-1} (A_{\rm b}\simeq \pot{2.8}{-8})$, the most probable solution was $n_{\rm b}\simeq 1.8$, $A_{\rm b}\simeq\pot{6}{-8}$, $y_{\rm re}\simeq\pot{3.8}{-7}$ and $\Delta^{\ast}\simeq\pot{9}{-9}$ for 10 times the sensitivity of {\it PIXIE}, illustrating the degeneracies of the parameter space, which can only be broken at very much higher experimental sensitivity. The distortion in this case is already dominated by a pure $\mu$-distortion, which contains too little information for constraining the two model parameters, making this behavior plausible.
In addition, this case is degenerate with the p-wave annihilation scenario discussed above, indicating that model independent constraints are hard to derive.

We also considered some cases with simultaneous step and change of the spectral index. Scenarios with spectral index $\nS'\simeq 1$ and large step amplitude are easiest to constrain, giving rise to a large overall signal and sufficient mixture of $\mu$-, $y$- and intermediate distortions. Similarly, models with bumps introduced by some scenarios with particle production \citep[e.g.,][]{Barnaby2010} could be directly constrained using spectral distortion; however, in this case degeneracies with decaying particle scenarios are expected. Given the large plausible parameter space, we stopped our discussion of this problem at this stage and look forward to more detailed investigations including realistic estimates of foregrounds and other instrumental effects as well as the combination with other cosmological data sets.

\section{Additional aspects}
In this section, we mention a few caveats that might affect the calculations carried out above at some significant level. In future work, these issues will have to be dealt with; however, they are beyond the scope of this paper.

\subsection{Cosmology dependence of the Green's function}
For the parameter estimations we assumed that the background cosmology does not affect the problem. This is true as long as effects (and errors) $\gtrsim 1\%$ are considered. However, for scenarios with large disparity of the associated signals (large p-wave annihilation with simultaneous small s-wave annihilation signal) or when differences of the signal with respect to a simple superposition of pure $\mu$- and $y$-distortions are important (e.g., when attempting to use the residual to learn something about the lifetime of a decaying particle), this assumption could lead to an underestimation of the errors, correlations and degeneracies.
In this situation, the computations carried out with the Green's function method have to be extended to include its cosmology dependence, and constraints should be derived simultaneously using CMB anisotropy, BAO, supernova and large-scale structure data. This in principle can be easily achieved; however, this will be left to future work.

\subsection{Sensitivity and frequency resolution}
In the previous sections, we only varied the sensitivity of the experiment, assuming that there is no fundamental limitation down to what precision foregrounds can be separated. Factors of a few improvements over {\it PIXIE} might be within reach, e.g., by extending the total time spent on spectral distortion measurements or by slightly improving the instrument; however, beyond that more sophisticated modeling and optimization will certainly be required. 
In this context, one can also vary the number of frequency channels. {\it PIXIE} is based on a Fourier-transform spectrometer (FTS) for which the spectral resolution is set by the dimension of the instrument or more specifically the maximal mirror excursion/stroke. Therefore, even improving the frequency resolution by a factor of a few might be challenging at this point and alternative approaches may have to be considered.
Adding more channels could generally help handling foregrounds and removal of narrow features [e.g., CO lines that pollute some of the {\it Planck} channels \citep{Planck2013components}], but since the primordial distortion signal is rather broad this does not lead to any obvious leverage with respect to the distortion signal itself. Sensitivity appears to be more crucial.

Another aspect is the lowest frequency channel. Residuals of the distortion with respect to a superposition of pure $\mu$- and $y$-distortions are noticeable even below the $30\,{\rm GHz}$ channel of {\it PIXIE} (see Fig.~\ref{fig:Decay_scenarios}), so that adding bands might further help to increase the sensitivity to different energy-release scenarios. In an FTS approach, this is again limited by the dimension of the instrument.
Ground-based observations might provide and interesting alternative avenue at those frequencies.

\subsection{Pre-recombination emission}
It was shown \citep{Chluba2008c} that the traces of neutral hydrogen and helium atoms present in the pre-recombination epoch respond strongly if the CMB spectrum is distorted. In this situation, cycles of uncompensated free-bound and bound-bound transitions lead to a significant enhancement of the cosmological recombination radiation, giving rise to additional spectral features that can be much more prominent than the lines produced during the respective recombination epochs. 
These features might allow us to tell the difference between post-recombination and pre-recombination $y$-distortions, thus providing another way to determine the time dependence of energy-release processes at $10^3 \lesssim z \lesssim 10^4$. 

Both at low and very high frequencies these features can become as large as the primordial spectral distortion itself \citep[see Fig. 12 of][]{Chluba2008c}. This therefore is an important correction that so far has not been taken into account consistently, but has the potential to alter the thermalization efficiency at $z\lesssim \pot{2}{6}$.
In particular, $\mu$-distortions are expected to cause strong induced atomic emission and thus enhance the atomic photon production rate at low frequencies significantly. This is because for a $y$-distortion, in the Rayleigh-Jeans part of the CMB spectrum, one has $\Delta I_\nu/I_\nu\simeq - 2 y$, which corresponds to a constant temperature shift. Resonances connecting different hydrogen levels at these frequencies are thus not strongly out of equilibrium among each other. For a $\mu$-distortion on the other hand, we have $\Delta I_\nu/I_\nu\simeq - \mu/x$, so that transitions at low frequencies are amplified due to the strong frequency dependence of the distortion.
Collisional processes (non-radiative) will diminish the efficiency of this process, but overall it could affect the detailed shape of the distortion caused by energy release at an important level.
Especially, if the primordial distortion is close to the {\it COBE}/{\rm FIRAS} limits, this problem will have to be solved if in the future high-precision spectral measurements will be used to learn details about the thermal history of our Universe.

\section{Conclusions}
\label{sec:conclusions}
We demonstrated that a {\it PIXIE}-type experiment will open an unexplored window to the early Universe. It will not only be able to rule out different scenarios with annihilating and decaying particles or large small-scale power, but could furthermore allow constraining specific model parameters on a case-by-case basis. 
Our analysis shows that improving existing limits on s-wave annihilation scenarios using spectral distortions will be challenging, but for p-wave annihilation with $\left<\sigma {\rm v}\right>\propto (1+z)$, accurate measurements (per cent precision) of the annihilation efficiency will be possible if $f_{\rm ann,p}\gtrsim 10^{-26}\,{\rm eV \, sec^{-1}}$ assuming {\it PIXIE}'s specifications (see Sect.~\ref{sec:annihil}). 
A p-wave annihilation signature could be detectable down to $f_{\rm ann,p}\simeq \pot{\rm few}{-28}\,{\rm eV \, sec^{-1}}$, a sensitivity that will be very hard to achieve using, e.g., CMB anisotropies.
Directly distinguishing the signatures from s- and p-wave annihilations, however, requires $\simeq 40$ times improvement of the experimental sensitivity over {\it PIXIE}, rendering an analysis along these lines more futuristic.
A combination of CMB anisotropy and spectral distortion measurements will help breaking the degeneracy of parameters in this case, providing additional leverage.

A {\it PIXIE}-type experiment will also be an exquisite probe for long-lived particles with lifetimes $t_{\rm X}\simeq \pot{6}{8}\,{\rm sec}-10^{10}\,{\rm sec}$ and decay efficiency, $f_{\rm X}/z_{\rm X}\gtrsim 1\,{\rm eV}$. In this case, distortions in principle can be used to constrain both the lifetime and decay efficiency with per cent precision. For particles with longer/shorter lifetime, the distortion signal is too close to a pure $y$/$\mu$-distortion to allow telling the particles lifetime. In both cases, however, tight limits on the abundance and mass of the particle can be derived from spectral distortion measurements. The extent to which different cases can be distinguished is a strong function of the sensitivity, so that pushing factors of a few improves matters significantly, complementing constraints derived from the CMB isotropies ($t_{\rm X}\simeq 380\,{\rm yrs}$) and the light-element abundances ($t_{\rm X}\simeq {\rm few \,minutes}$).

Finally, CMB spectral distortion can also be used to constrain different early-universe models, responsible for primordial perturbations of the cosmic fluid at wavenumbers $1\,\Mpc^{-1}\lesssim k \lesssim \pot{\rm few}{4}\,\Mpc^{-1}$. Extrapolating the power spectrum from large scales, $k\lesssim 1\,\Mpc^{-1}$, all the way to $k\simeq \pot{\rm few}{4}$, we find that improvements by a factor of  $\simeq 100- 200$ in the spectral sensitivity of {\it PIXIE} are needed to deliver an independent probe that is competitive with current and upcoming CMB anisotropy data; however, the parameter space in principle is wide open, and stringent limits on a model-by-model basis can be obtained already with a {\it PIXIE}-type experiment. These limits remain rather model dependent even for much higher sensitivity, and multimodal solutions are obtained.

Simple models with a step in the small-scale power spectrum can be constrained very well, if the step appears at $k_{\rm s}\simeq 20\,\Mpc^{-1}-50\,\Mpc^{-1}$ giving $\Delta k_{\rm s}/k_{\rm s}\lesssim 6\%$ and $\Delta\Delta A_\zeta/\Delta A_\zeta\lesssim 2\%$ for step amplitudes $\Delta A_\zeta\gtrsim \pot{4}{-7}$. Similar to the decaying particle scenarios, we find that these statements are a strong function of the experimental sensitivity, thus factors of a few improvement provide a large leverage. Constraints on specific parameters of models with flaring small-scale power are hard to obtain, even at much higher sensitivity. Still, CMB spectral distortions translate into tight integral constraints on the inflaton's trajectory.

Given the huge possible parameter space, a {\it PIXIE}-type experiment provides a unique opportunity to probe early-universe models and particle physics, no matter if a primordial distortion signal is detected and can be linked to a specific model or not. We look forward to improving our estimates, accounting for more realistic foregrounds and other experimental aspects, as well as combined constraints with other cosmological data sets.

\small

\section*{Acknowledgements}
JC thanks Yacine Ali-Ha{\"i}moud, Stefan Hilbert, Marc Kamionkowski, Rishi Khatri, Eiichiro Komatsu, Brice Menard, Jacqueline Radigan, Rashid Sunyaev, Eric Switzer, and Geoff Vasil for useful discussions and comments on the manuscript. 
Use of the GPC supercomputer at the SciNet HPC Consortium is acknowledged. SciNet is funded by: the Canada Foundation for Innovation under the auspices of Compute Canada; the Government of Ontario; Ontario Research Fund - Research Excellence; and the University of Toronto. This work was supported by the grants DoE SC-0008108 and NASA NNX12AE86G.

\vspace{-3mm}

\bibliographystyle{mn2e}
\bibliography{Lit}

\end{document}